\documentclass[final, 5p, times, twocolumn]{elsarticle}

\usepackage{float}
\usepackage{graphicx} 
\usepackage{enumitem}
\usepackage{booktabs}

\usepackage{amsmath}
\usepackage{hyperref}
\hypersetup{
  colorlinks=true, 
  linkcolor=blue, 
  urlcolor=blue, 
  filecolor=blue, 
  citecolor=blue 
}

\usepackage{float}
\usepackage{siunitx}
\usepackage{chemformula}

\usepackage{xcolor}
\usepackage{xurl} 
\usepackage{svg}

\usepackage[version=4]{mhchem}
\usepackage{subcaption}

\usepackage[acronym]{glossaries}

\newacronym{ai}{AI}{Artificial Intelligence}
\newacronym{api}{API}{Application Programming Interface}
\newacronym{aware}{AWARE}{Available Water Remaining}
\newacronym{cee}{CEE}{CPU Energy Efficiency}
\newacronym{cf}{CF}{Carbon Footprint}
\newacronym{dc}{DC}{Data center}
\newacronym{entso}{ENTSO-E}{European Network of Transmission System Operators for Electricity}
\newacronym{es}{ES}{Environmental Score}
\newacronym{gd}{GD}{Green Director}
\newacronym{gee}{GEE}{GPU Energy Efficiency}
\newacronym{ghg}{GHG}{Greenhouse Gas}
\newacronym{gpu}{GPU}{Graphical Processing Unit}
\newacronym{gs}{GS}{Green Score}
\newacronym{hee}{HEE}{Hardware Energy Efficiency}
\newacronym{jrc}{JRC}{Joint Research Center}
\newacronym{ict}{ICT}{Information and Computing Technologies}
\newacronym{lca}{LCA}{Life-cycle assessment}
\newacronym{pue}{PUE}{Power Usage Effectiveness}
\newacronym{tdp}{TDP}{Thermal Design Power}
\newacronym{vo}{VO}{Virtual Organization}
\newacronym{wf}{WF}{Water Footprint}
\newacronym{wi}{WI}{Water Impact}
\newacronym{wms}{WMS}{Workload Management System}
\newacronym{wue}{WUE}{Water Usage Effectiveness}
\newacronym{wulca}{WULCA}{Water Use in Life Cycle Assessment}
\newacronym[
  longplural={GreenSiteDirectors},
  shortplural={GSDs}
]{gsd}{GSD}{GreenSiteDirector}

\journal{Future Gener. Comput. Syst.}

\begin{document}

\begin{frontmatter}



    \title{GreenDirector: carbon- and water-aware workload placement for sustainable computing}

    \author[ifca]{Jaime Iglesias Blanco}
    \author[ifca]{Ignacio Heredia}
    \author[ifca]{María Castrillo}
    \author[cppm]{Andrei Tsaregorodtsev}
    \author[cppm]{Mazen Ezzeddine}
    \author[ifca]{Álvaro López García\corref{cor1}}
    \ead{aloga@ifca.unican.es}
    \cortext[cor1]{Corresponding author}

    \affiliation[ifca]{organization={Instituto de Física de Cantabria (IFCA), CSIC-UC},
        addressline={Avda. los Castros s/n},
        postcode={39005}, 
        city={Santander}, 
        state={Cantabria}, 
        country={Spain}
        }
    \affiliation[cppm]{
        organization={Aix Marseille Univ, CNRS/IN2P3, CPPM},
        postcode={13288},
        city={Marseille},
        country={France}
    }

    \begin{abstract}
    The rapid growth of data center electricity demand, accelerated by AI, makes carbon-only accounting an incomplete measure of computing's environmental impact: low-carbon electricity mixes are often water-intensive, and the resulting harm depends on local, seasonal scarcity rather than on the volume of water consumed. We propose the Environmental Score (ES), a unified, dimensionless index in $[0, 100]$ that jointly captures the carbon footprint and the spatial-temporal, scarcity-weighted water impact of the electricity a workload consumes. It combines real-time, cross-border electricity flow tracing with monthly AWARE2.0 water-scarcity characterization factors, weighting global greenhouse-gas emissions together with the local, seasonal severity of water stress. Building on it, we define the Green Score (GS), a scheduling metric proportional to the useful computational work delivered per unit of real environmental impact, which also accounts for data center power and hardware efficiency. We add both metrics as a green-affinity feature to the GreenDirector schedulers of two production federated infrastructures, the AI4EOSC scientific cloud and the DIRAC workload management system. In AI4EOSC, a cluster-filling experiment over four pan-European providers shows that greener sites are filled first without degrading scheduling latency or end-user experience. In DIRAC, a trace-driven simulation of 133{,}631 jobs and a preliminary production deployment for the KM3NeT community reduce carbon emissions and improve carbon efficiency by about 40\%, while making the carbon-water trade-off explicit when the lowest-carbon site also carries higher water stress. The results show that hydrological stress can be dynamically weighted into workload placement in live, multi-tenant systems.
    \end{abstract}
    \begin{keyword}
    Sustainable computing \sep Carbon-aware scheduling \sep Water scarcity footprint \sep Water-energy nexus \sep Federated cloud computing \sep Grid computing
    \end{keyword}
\end{frontmatter}

\section{Introduction}
\label{sec:introduction}

Electricity consumption in the computing sector has grown notably in recent years. In particular, according to recent industry assessments \cite{shehabi2024datacenter}, the rapid expansion of the data center industry is creating an unprecedented energy demand. Given the critical importance of water use concerns in modern data centers ---particularly with the advent of \gls{ai}---, focusing solely on \gls{ghg} emissions provides an incomplete picture of a data center's ecological footprint and environmental impact.

During the operational phase (excluding the upstream or downstream footprint during the entire life cycle), data centers utilize water in two distinct ways:
\begin{itemize}
    \item \textbf{On-site water withdrawal} occurs in the data center facility itself, normally for cooling purposes. This aspect of water use is highly visible, and given its evident impact \cite{lei_water_2025, li_making_2025}, it can lead to social tensions in locations experiencing water scarcity \cite{ngata_cloud_2025, privette_data_2026, diaz_bejarano_thirsty_2026, shah_four_2026}. The rise of high-density computing and \gls{ai} has exacerbated this issue, as modern \glspl{gpu} generate such intense heat that cooling relies heavily on local water resources. 
    \item \textbf{Indirect off-site use}, which refers to the water consumed to generate the electricity used by a facility, where the \gls{cf} and \gls{wf} of electricity generation are strongly interconnected and cannot be analysed independently.
\end{itemize}

Regarding the former, evaluating \gls{cf} and \gls{wf} independently is insufficient due to the inherent trade-offs within different energy mixes. For example, a region might offer a low-carbon energy mix that is water-intensive at the same time, which could be even worse if the region suffers from water scarcity \cite{lohrmann_troubled_2023}. This makes a decarbonization strategy that ignores the water-energy nexus risk shifting the environmental burden from global climate systems to local hydrological basins \cite{JRC_water_energy} due to the intensive use of water that some low-carbon energy systems do, especially biomass and hydropower \cite{vanham_consumptive_2019}. Transitioning to round-the-clock renewable energy availability ---the concept of providing a continuous supply of renewable energy, 24 hours a day, 7 days a week--- needs storage solutions to provide generation flexibility (such as hydropower), which can have a strong \gls{wf}. This emphasizes the need to address the use and management of energy and water resources simultaneously, trying to maximize opportunities in both systems, as optimizing both the carbon and the \gls{wf} at the same time with the available technologies is definitely challenging.

To achieve true environmental sustainability, we argue that it is imperative to move beyond single-dimensional reporting and integrate water consumption (and other impact signals) into the decision-making framework. Regarding water consumption, it is important not only how much water is consumed, but also how much water stress it generates. While the \gls{ghg} emissions impact, at least in terms of climate change effects, is not dependent on the place where they are generated, the ISO 14046:2014 \cite{iso14046} standard stipulates that a \gls{wf} is not merely a volume of water, but must include an assessment of its environmental significance based on local impact and vulnerability.

In this work, we integrate multiple environmental footprint and impact signals into the scheduling decisions of distributed computing infrastructures, taking into account the spatial-temporal dimensions of water scarcity. The specific contributions are the following:

\begin{itemize}
    \item \textbf{The \acrfull{es}}: a unified, dimensionless index ($[0, 100]$) that combines the \gls{cf} and the spatial-temporal, scarcity-weighted \gls{wi} of the electricity a workload consumes. Unlike traditional ``carbon-only'' metrics, it is derived from real-time, cross-border electricity flow tracing and monthly AWARE2.0 characterisation factors, in line with ISO 14046, and is designed to be extended with further impact categories without changing the components that consume it.
    \item \textbf{The \acrfull{gs}}: a scheduling metric built on the \gls{es} that expresses the useful computational work delivered per unit of real environmental impact. It incorporates the data center \gls{pue} and a per-node hardware-efficiency term, and is instantiated per workload profile (CPU-only or CPU+GPU).
    \item \textbf{An implementation of both metrics as a green-affinity feature} of the \glspl{gd} of two production federated infrastructures: the AI4EOSC platform, deployed within the EOSC Ecosystem over cloud computing infrastructures~\cite{HEREDIA2026108672}, and the DIRAC Interware~\cite{stagni_dirac_2020} \gls{wms} (grid), leaving the existing scheduling logic otherwise unchanged.
    \item \textbf{An evaluation} comprising a spatial-temporal environmental characterisation of nine European infrastructure providers; a score-inspection and cluster-filling validation in AI4EOSC, where greener providers are filled first without degrading scheduling latency or end-user experience; and, in DIRAC, a trace-driven simulation of 133{,}631 jobs together with a preliminary production deployment for the KM3NeT community, which reduce carbon emissions by 42.8\% and improve carbon efficiency by 49.5\%, respectively, while making the carbon-water trade-off explicit.
\end{itemize}

The remainder of the manuscript is structured as follows:
Section~\ref{sec:related} introduces the related work in the area, both from the sustainability and the \gls{ict} sides; Section~\ref{sec:methodology} outlines the methodology that we have followed to develop the \gls{es} and \gls{gs} metric; Section~\ref{sec:implementation} introduces how we have implemented the \gls{es} as a practical example in the AI4EOSC's and DIRAC's \glspl{gd}; Section~\ref{sec:results} provides the evaluation results and discussion. Finally, conclusions are drawn out in Section~\ref{sec:conclusion}.

\section{Related work}
\label{sec:related}

\subsection{From carbon-aware to water-aware computing}
The mainstream of efforts to mitigate the environmental impact of \gls{ict} (and in particular, cloud computing) has focused on energy efficiency and reducing \gls{ghg} emissions. Beyond the optimization of workloads themselves (i.e., making the software more efficient), there is a large body of work that tackles different spatial-temporal optimization methods on scheduling frameworks to align the workload and compute tasks execution with locations (spatial dimension) or periods (temporal dimension) where their \ch{CO2} emissions are lower. In \cite{zanotto_user-centric_2025}, the authors implement a carbon-intensity forecaster that they use to dynamically adjust workload executions on top of Kubernetes clusters. In \cite{souza_casper_2024}, the authors design a carbon-aware scheduling and provisioning system that primarily minimizes the carbon footprint of distributed web services. Similarly, \cite{schweisgut_carbon-aware_2025} focuses on scientific computing, minimizing carbon emissions on a parallel computing platform with a time-varying mixed (renewable and non-renewable) energy supply. The study in \cite{bostandoost_data-driven_2025} presents a carbon-aware scheduling algorithm, focusing on reducing carbon emissions of delay-tolerant batch workloads. Shaoshu et al. \cite{zhu_does_2026} discuss whether energy efficiency implies carbon awareness. They utilize AI workloads and carbon-intensity traces showing that energy-efficient scheduling does not necessarily yield lower carbon emissions, as spatial–temporal variability and hardware heterogeneity introduce non-linear trade-offs.

However, the water-energy nexus is often overlooked, although the sector's water consumption is nowadays emerging as a critical sustainability bottleneck. As already introduced, recent works emphasize that data centers consume water in two distinct ways: on-site facility cooling and off-site electricity generation \cite{lei_water_2025, privette_data_2026, shah_four_2026}. This dual consumption can trigger severe localized resource conflicts, particularly when facilities operate in drought-prone or water-stressed regions, such as Mediterranean regions \cite{privette_data_2026, diaz_bejarano_thirsty_2026, shah_four_2026}. To better understand these localized impacts, Jiang et al. \cite{jiang_facility-level_2025} conducted a bottom-up analysis of facility-level, energy-driven \glspl{wf} across Chinese data centers. Their scenario-based projections point to the critical necessity of evaluating localized water scarcity rather than relying on absolute volumetric consumption.
In \cite{samanta_water_2026}, the authors highlight the need for methodology standardization for water-aware operations in data centers, revealing promising opportunities for water-aware scheduling considering regional water variations and life-cycle impacts. More specifically on AI workloads, \cite{moore_sustainable_2025} introduces a carbon- and water-efficient Large Language Model scheduling mechanism across geographically distributed resources.


\subsection{Multi-dimensional environmental optimization}

Recognizing the limitations of single-metric optimization, recent studies have started to explore multi-dimensional sustainability frameworks. As already mentioned, a clear challenge in this domain is the inherent water-energy nexus \cite{JRC_water_energy}. Jiang et al. introduce in \cite{jiang_waterwise_2025} their \textit{WaterWise} scheduling framework. In their work, they show that optimizing exclusively for low carbon intensity can inadvertently penalize a facility's \gls{wf}, as certain low-carbon power sources (most notably hydropower, biomass, and nuclear energy) require substantial operational water withdrawal and evaporation during electricity generation.

Expanding the optimization space further, the work of \cite{attenni_spatio-temporal_2025} implements a multi-dimensional sustainability-aware cloud computing scheduling approach, including not only carbon and water, but also land use impacts. They conclude that both spatial and temporal shifting can significantly reduce the environmental footprint of cloud workloads in terms of carbon emissions, water consumption, and land use. In our opinion, while allocating the workloads from an ecological perspective can truly reduce the carbon and \gls{wf}, making a more intensive use of low-carbon energy and reducing the demand for high-carbon energy systems, the land use will likely remain constant. The space occupied by energy generation infrastructure is not dependent on the instant energy generation, while carbon emissions and water consumption are. In \cite{noauthor_sustainability-aware_nodate}, Hoffman and Majuntke introduce ORCA, a sustainability-aware workload shifting framework that considers global climate impacts and heterogeneous local criteria, integrating different region- and time-dependent signals for the impact characterization.

\subsection{Unified spatial-temporal metrics}

While existing works successfully highlight the necessity of co-optimizing carbon and water (and other signals), there is a lack of practical implementation that unifies these signals into an actionable, standardized operational metric. Most current approaches either treat \gls{wf}s as static annual averages or fail to incorporate standardized \gls{lca} scarcity weightings (such as \gls{aware}2.0) into real-time job placement algorithms. 

Our work bridges this gap by introducing the \gls{es}, demonstrated with a practical implementation within the AI4EOSC's and DIRAC's \glspl{gd} and their associated \gls{gs} scheduling metric. Unlike previous frameworks, our approach dynamically synthesizes real-time, cross-border electricity tracking (capturing energy imports and exports via Wattnet \cite{castrillo_melguizo_wattnet_2026}) with seasonal, localized water stress characterization factors. By translating complex, multi-dimensional \gls{lca} impacts into an intuitive, dimensionless affinity score ($[0, 100]$), we enable federated cloud orchestration systems (such as Nomad) to actively co-optimize carbon and water impacts in a live production environment.

\section{Methodology}
\label{sec:methodology}

In this section, we first describe how we combine different environmental footprints and impacts into a single metric that we define as \acrfull{es} (Section \ref{sec:score}). Secondly, we introduce the \acrfull{gs} as a new metric to better prioritize ``greener'' (in terms of impact) \glspl{dc} when performing scheduling decisions.

\subsection{Impact metric calculation}
\label{sec:score}

Taking into account an \gls{ict} perspective where a given workload is to be executed at a given location, we consider that its ecological component is based on the \gls{cf} and \gls{wf} of the electricity consumed by the execution of such workload. That is to say that the environmental impact considered here corresponds exclusively to the resources consumed in the generation of the electricity used to operate the \gls{dc}, analogous to a Scope 2 emission in \gls{cf} accounting, and does not include on-site resource consumption such as the water used for cooling purposes. Data about the footprints are obtained from Wattnet, which is an online service\footnote{\url{https://wattnet.eu/}} to track the environmental footprint of electricity across Europe. Wattnet implements a flow tracing algorithm and jointly assesses the \gls{cf} and \gls{wf} of electricity production. It dynamically calculates both \gls{cf} and \gls{wf} with high temporal resolution, considering not only each region's energy generation mix but also energy flows due to imports and exports between neighbouring regions in Europe with a high temporal resolution (fifteen minutes). It mainly uses the data provided by the \gls{entso} Transparency Platform, but also includes other relevant data sources.

However, when allocating workloads across data centers, considering their \gls{cf} and \gls{wf} must account for their real impact rather than their absolute values. In this regard, one point to bear in mind is that the impact of these two footprints differs not only in their dimensions and scales but also in their geographical and seasonal dependence. Unlike water consumption, the impact of carbon emissions is both location- and season-independent. Although it is known that some regions suffer more from climate change than others, the Global Warming Potential of a gram of $\ce{\ch{CO2}}$ emitted in a certain location has the same physical effect on global climate as a gram emitted in any other location, regardless of the season of the year. Therefore, the \gls{cf} value can be used as a direct proxy of carbon impact. Wattnet differentiates between \textit{local} and \textit{global} \gls{cf} to refer to the \gls{cf} of the electricity generated in the country or region in question, while the \gls{cf} considers the successive mixing due to imports and exports, respectively. The appropriate value to calculate the carbon impact is the global \gls{cf}. 

Regarding water, considered a renewable resource at the global scale, its consumption has local impacts that are geographically and seasonally dependent. The same volume of water withdrawn in a water-abundant region produces a negligible environmental impact, whereas identical consumption in a water-scarce region may cause severe ecological stress. Furthermore, water availability varies across seasons: in Mediterranean and semi-arid climates, a withdrawal that is inconsequential in winter may place significant pressure on local water resources during the dry summer months. 

In this sense, we have extended Wattnet's API to also provide other data beyond the direct footprints, namely the water impact (described in Section~\ref{sec:water_impact}) and the new unified environmental score (Section~\ref{sec:es}) that we have named \acrfull{es}.

\subsubsection{Water impact}
\label{sec:water_impact}

To account for the geographical and seasonal dependency of the \gls{wi}, it is calculated in accordance with the guidelines of ISO 14046, using the updated and improved method for water scarcity impact assessment in \gls{lca}, \gls{aware}2.0~\cite{seitfudem_updated_2025, georg_seitfudem_2025_16332127}, which provides characterization factors at both monthly and annual resolution. Spatially, \gls{aware} factors are defined at the country level and further disaggregated into multiple subnational regions. This approach allows for a more precise capture of local variability in water availability, consistent with the zoning structure used in Wattnet, largely determined by the granularity of electricity transmission data. \gls{aware} is the consensus method for assessing water scarcity impact per unit of water consumed, recommended by the \gls{wulca} initiative and adopted in the European Commission's Product Environmental Footprint (PEF) method~\cite{boulay_wulca_2018, seitfudem_updated_2025, georg_seitfudem_2025_16332127}.

\begin{figure*}[htbp]
    \centering
    \includegraphics[width=\textwidth]{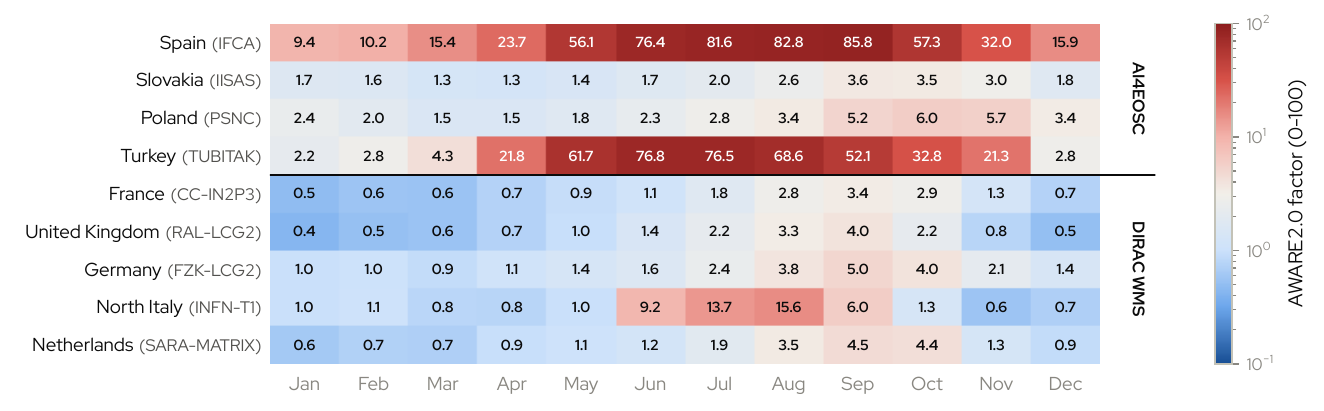}
    \caption{Monthly \gls{aware}2.0 characterisation factors across the paper's two use cases: the four AI4EOSC data centers (IFCA/Spain, IISAS/Slovakia, PSNC/Poland, TUBITAK/Turkey) and a selection of DIRAC Workload Management System sites, one per country (CC-IN2P3/France, RAL-LCG2/United Kingdom, FZK-LCG2/Germany, INFN-T1/North Italy, SARA-MATRIX/Netherlands). IFCA and TUBITAK exhibit pronounced summer peaks ($>$70) driven by Mediterranean water scarcity, while IISAS and PSNC remain consistently low ($<$6) throughout the year. Most DIRAC WMS sites stay comparably low year-round ($<$5), except INFN-T1, whose North Italy zone shows a moderate summer peak ($\sim$15).}
    \label{fig:aware_heatmap}
\end{figure*}

As shown in Figure~\ref{fig:aware_heatmap}, in the case of AI4EOSC data centers, the \gls{aware}2.0 factor for IFCA and TUBITAK varies by more than an order of magnitude between winter and summer, whereas IISAS and PSNC show almost no seasonal variation. This empirically confirms the need to use monthly factors (Equation~\ref{WI}) rather than a fixed annual value, particularly for Mediterranean zones, where ignoring seasonality would drastically underestimate water impact during the summer months. The same pattern holds for the DIRAC Workload Management System sites included in the same figure: CC-IN2P3, RAL-LCG2, FZK-LCG2 and SARA-MATRIX remain consistently low throughout the year ($<$5), while INFN-T1, located in North Italy, shows a distinct summer increase (up to $\sim$15 in July–August). This confirms that pronounced seasonal water-stress variation is not specific to AI4EOSC's Mediterranean deployments, but a broader feature of European electricity zones that any multi-site infrastructure, including DIRAC WMS, must account for when estimating water impact.

Due to energy imports and exports, the energy consumed in a given country or region is produced not only in that geographical location, but also in neighbouring countries and their neighbours, and so on. Similarly, the \gls{wi} of the energy consumed by a data center is distributed among all these countries or regions in proportion to their share of the energy mix of the country or region in question. The value of these shares is obtained through Wattnet, which implements the methodology described in \cite{castrillo_melguizo_wattnet_2026}. Then, for each generation source $k$ contributing to the mix of zone $i$, its local \gls{wf} $WF_k$ is weighted by its share $P_{i,k}$ of the $k$ zone mix in the $i$ zone mix and by the corresponding \gls{aware} characterisation factor in the temporal context. The global \gls{wi} of zone $i$ is then calculated as expressed in Equation~\ref{WI}:

\begin{equation}\label{WI}
    WI_i(t) = \sum_{k} P_{i,k}(t) \cdot WF_k \cdot AWARE_k(t)
\end{equation}

where $P_{i,k}(t)$ is the share of zone $k$ in the electricity mix of zone $i$ at time $t$, $WF_k$ is the \gls{wf} intensity of zone $k$ (\si{l\per\kWh}), and $AWARE_k(t)$ is the \gls{aware}2.0 characterisation factor for zone $k$ at time $t$. The spatial-temporal resolution of the \gls{aware} factor can only be meaningfully applied to the \textit{operational} scope, since the location and time of electricity generation are known, and therefore so is the region where the associated water-stress impact occurs; in this case, the monthly factor corresponding to the month of $t$ and the generation zone is used, capturing seasonal variability in water availability. For the remaining life-cycle stages, however, neither the location nor the timing of the underlying processes can be determined with the same precision, since they depend on decisions made by third parties outside the scope of this study (e.g., component manufacturing or plant decommissioning). Consequently, a spatially and temporally resolved \gls{aware} factor cannot be applied to the \textit{life-cycle} scope.

\subsubsection{Unified environmental score}
\label{sec:es}

Unifying two metrics, such as the operational \gls{cf} and \gls{wf}, involves normalising and weighting them. Normalisation allows scaling the data so that it can be compared regardless of
its original units, while weighting allows giving different importance to each of the impacts. 

As a normalization technique, ratio scaling has been used (Equation~\ref{scaling}). It transforms each zone's impact into a unitless value in the range $(0,1]$, scaled relative to the most polluting zone available at that instant.

\begin{equation}\label{scaling}
X^{r}(t) = \frac{X(t)}{X_{max}(t)}
\end{equation}

As in the previous formulation, $X_{max}(t)$ is computed at each time interval over the full set of European zones available at that instant, so $X^{r}(t)=1$ corresponds to the most polluting zone (highest impact) at that instant. Unlike min-max scaling, ratio scaling does not force the cleanest zone in the comparison set to a value of exactly zero: since real environmental impact is never physically zero, this behaviour is more conceptually consistent with the underlying quantities being represented, at the cost of the scaled values no longer spanning the full $[0,1]$ interval.

The \gls{cf} weight ($W_{C}$) and the \gls{wi} weight ($W_{W}$) are derived from the \gls{jrc} report~\cite{JRC_LCA_2018} on the development of weightings for calculating the environmental footprint. In the report, which includes 16 impact categories, climate change accounts for approximately 21\% and water for 8.5\%. Normalising these two categories to sum to unity yields $W_{C} = 0.71$ for the \gls{cf} and $W_{W} = 0.29$ for \gls{wi}. We apply these as fixed relative-importance weights to the zone-relative, ratio-scaled indicators defined in Equation~\ref{scaling}, rather than attempting to reproduce the JRC's own person-equivalent normalisation basis; as a result, the realised average contribution of \gls{wi} to the combined score can be smaller than 29\% in practice, since $WI^{r}$ is empirically concentrated near zero for most European zones (Section~\ref{sec:water_impact}, Figure~\ref{fig:aware_heatmap}) while $CF^{r}$ is more evenly distributed. We deliberately do not introduce a corrective rebalancing factor to force the realised average ratio to match 71/29, as doing so would make the weighting depend on the arbitrary composition of the zone set under comparison at a given time, rather than remaining a stable, reproducible parameter.


The \gls{cf} and \gls{wi} weights are combined directly into the \gls{es}, as shown in Equation~\ref{eq:es}:

\begin{equation}\label{eq:es}
\mathrm{ES}_i(t) = 100\left[1-\left(W_C\,CF_i^{r}(t) + W_W\,WI_i^{r}(t)\right)\right]
\end{equation}

Since $CF_i^{r}(t), WI_i^{r}(t) \in (0,1]$ and $W_C + W_W = 1$, the weighted sum inside the brackets is itself bounded in $(0,1]$ by construction, which keeps $\mathrm{ES}_i(t)$ within the intended $[0,100]$ range.

The inversion applied in Equation~\ref{eq:es} ensures that the resulting score is directly interpretable as an environmental quality index in the range $[0, 100]$: a score of 100 corresponds to the cleanest zone in the current interval, while a score of 0 corresponds to the most polluting one. Without this inversion, the weighted sum of the scaled impacts would yield the opposite behaviour, with higher values indicating greater environmental burden. This convention is consistent with using the score as an affinity metric in \gls{gd}: a higher score translates directly into a stronger preference for placing workloads in that zone, without requiring any additional sign reversal downstream in the scheduling logic. The same bracketed quantity is used directly in Section~\ref{sec:green score} to define \gls{gs}, via $1-\mathrm{ES}_i(t)/100$.

\begin{figure*}[htbp]
    \centering
    \includegraphics[width=\linewidth]{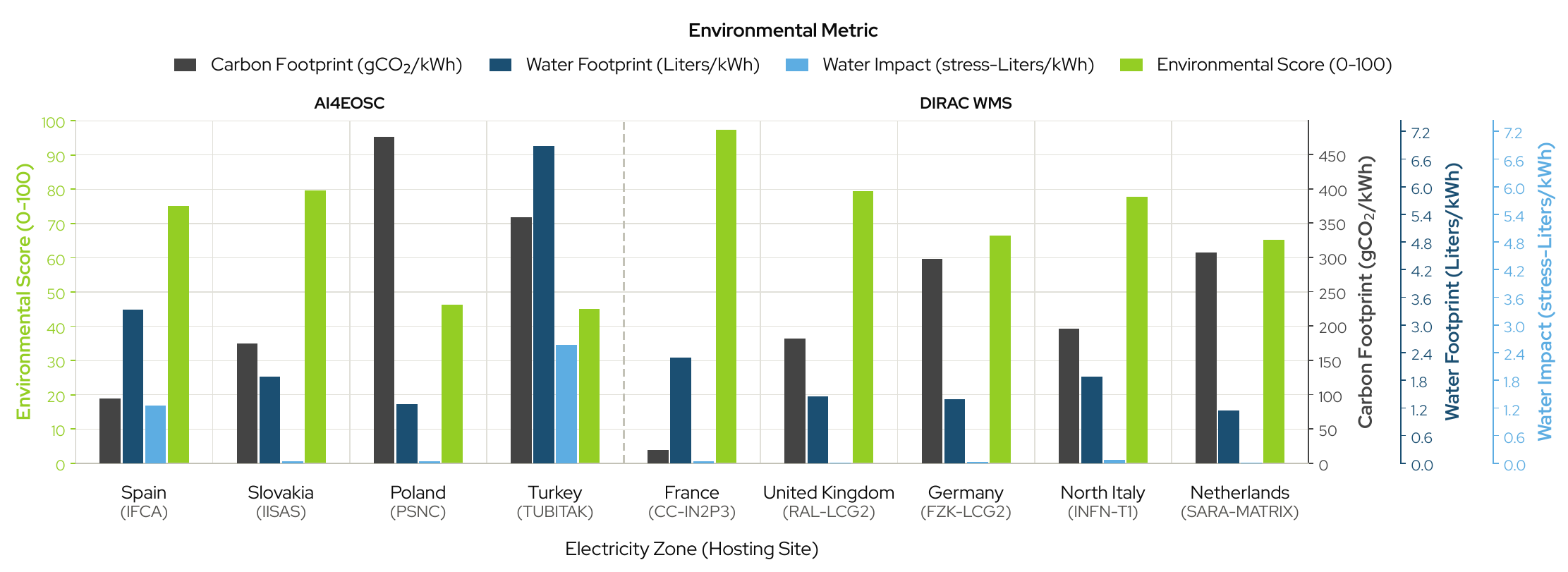}
    \caption{Annual average \gls{cf}, \gls{wf}, \gls{wi} and \gls{es} across the paper's two use cases: the four AI4EOSC platform infrastructure providers (IFCA/Spain, IISAS/Slovakia, PSNC/Poland, TUBITAK/Turkey) and a selection of DIRAC Workload Management System sites, one per country (CC-IN2P3/France, RAL-LCG2/United Kingdom, FZK-LCG2/Germany, INFN-T1/North Italy, SARA-MATRIX/Netherlands), computed from Wattnet operational data at 15-minute resolution over the period July 2025–June 2026, including cross-border electricity exchange (global coverage).}
    \label{fig:scores_vs_countries}
\end{figure*}

Finally, Figure~\ref{fig:scores_vs_countries} shows how the \gls{cf} and \gls{wi} are combined into the overall \gls{es} for several AI4EOSC and DIRAC infrastructure providers, as given by Equation~\ref{eq:es}. Among the AI4EOSC providers, IFCA and IISAS benefit from low-carbon energy mixes and obtain the highest scores; IFCA does so despite a non-negligible \gls{wi} (annual average 1.25~stress-L/kWh), whereas IISAS's \gls{wi} stays close to zero (0.05~stress-L/kWh). In contrast, TUBITAK and PSNC are penalised for structurally different reasons. TUBITAK's pronounced \gls{wi} is explained by Turkey's electricity mix: despite its significant solar potential, hydroelectric power remains the country's dominant generation source, and Turkey holds one of the largest hydroelectric capacities in Europe. This reliance on hydropower, rather than solar, is what drives the elevated \gls{wi} observed for this provider, as reflected in the corresponding bars in the figure.

PSNC, on the other hand, is penalised primarily by high carbon intensity rather than by water-related factors. It is worth emphasising that the \gls{es} does not depend on raw water usage but on \gls{wi}, a distinction that accounts for local water stress, and which is explicitly represented as a separate series in the figure. As a result, a provider with high carbon emissions but low \gls{wi} (as is the case for PSNC) can still obtain a comparatively favourable \gls{es}, which may initially seem counterintuitive if one considers carbon emissions alone.

The same figure also includes the selected DIRAC Workload Management System sites. Unlike IFCA or TUBITAK, none of these five zones lies under meaningful water stress: their AWARE2.0-weighted \gls{wi} stays close to zero throughout (0.02--0.07~stress-L/kWh), so their \gls{es} is driven almost entirely by carbon intensity. CC-IN2P3 benefits from France's nuclear-dominated, low-carbon grid (19.4~gCO$_2$/kWh) combined with negligible \gls{wi} (0.04~stress-L/kWh), yielding the highest \gls{es} of all nine providers considered (97.4). RAL-LCG2 and INFN-T1 occupy an intermediate position, with moderate carbon intensities (181.3 and 195.7~gCO$_2$/kWh, respectively) offset by comparably negligible \gls{wi}, resulting in favourable scores (79.6 and 77.8). FZK-LCG2 and SARA-MATRIX illustrate the same dynamic observed for PSNC: despite comparatively high carbon footprints (298.2 and 307.6~gCO$_2$/kWh, respectively, the two highest among the DIRAC WMS sites), their negligible \gls{wi} (0.02~stress-L/kWh in both cases) keeps their \gls{es} (66.6 and 65.3) from falling as sharply as a carbon-only metric would suggest, reinforcing that carbon intensity and \gls{wi} are independent, decoupled dimensions of environmental performance.

\subsection{GreenScore}
\label{sec:green score}

While \gls{es} (Section~\ref{sec:score}) provides a bounded index suited to monitoring and comparing zones, scheduling decisions require a quantity that scales proportionally with real environmental impact rather than a relative ranking of it. We therefore build \gls{gs} on the proportional, unbounded quantity $1-\mathrm{ES}_i(t)/100$ (Equation~\ref{eq:es}) rather than on \gls{es} itself.

To characterize the amount of useful computational performance that is delivered per unit of power consumption, we firstly define the \gls{cee}, as shown in Equation~\ref {CEE}:

\begin{equation}\label{CEE}
   CEE = \frac{\mathit{CPUNormalizationFactor} \cdot \mathit{ncores}}{\mathit{TDP}}
\end{equation}

where $CPUNormalizationFactor$ is a single-core performance scaling factor (typically derived from a benchmark) that normalises CPU performance across heterogeneous hardware. It reflects the relative number of instructions
or computational work that can be executed per unit time compared to a reference CPU; and the \gls{tdp}, which is the maximum power (in Watts) that the CPU is designed to dissipate under sustained load, serving as a proxy for the energy consumption of the processor during execution.

The \gls{cee} metric provides a hardware-level efficiency indicator, capturing how effectively a given CPU server converts electrical power into useful computation. Higher \gls{cee} values correspond to more energy-efficient execution platforms, making them preferable candidates for green-aware scheduling. \gls{cee} can be aggregated per \gls{dc} and/or per time window to obtain stable efficiency indicators for scheduling. When node-level scheduling is available, using disaggregated CEE values is preferred for more precise scheduling.

In the case of AI4EOSC, the platform is heavily oriented to manage AI training workloads, thus heavily reliant on \glspl{gpu}. Therefore, we can extend the CEE to introduce a \gls{gee}:

\begin{equation}\label{GEE}
    GEE = \frac{TFLOPs}{TDP}
\end{equation}

which, in the case of AI training workloads (that make use of both CPUs and \gls{gpu}s), can be combined into an overall \gls{hee}:

\begin{equation}\label{HEE}
    HEE = \alpha_C \cdot CEE_{n} + \alpha_G \cdot GEE_{n}
\end{equation}

where $_n$ means that each efficiency has been normalized to a common scale so that they are comparable, using the same ratio scaling of Equation~\ref{scaling} (each value divided by the maximum across the providers eligible for that workload profile), and the $\alpha$ factors account for the relative amount of energy consumed by either the CPU or the \gls{gpu} in a typical AI workload. Initial estimates ($\alpha_C=0.2$ and $\alpha_G=0.8$) will be further refined based on the monitoring of future workloads.

Equation~\ref{HEE} is instantiated per workload profile, because each profile has a different set of candidate providers. CPU-only workloads can run on all four AI4EOSC providers; for them $\alpha_C=1$ and $\alpha_G=0$, so the hardware term reduces to $CEE_{n}$ with the maximum taken over the four providers. CPU+GPU workloads can only run on the \gls{gpu}-equipped providers (here IFCA and PSNC); for them the full \gls{hee} applies, with $CEE_{n}$ and $GEE_{n}$ each scaled by the maximum over that \gls{gpu}-equipped subset. In both cases, $CEE_i$ (and $GEE_i$) is the arithmetic mean of the per-model CPU (respectively \gls{gpu}) efficiencies of provider $i$, assuming an equal node count per model; IISAS, for which no per-model CPU data is available, uses the fleet-average \gls{tdp} and normalization factor.

\begin{figure*}
    \centering
    \includegraphics[width=\textwidth]{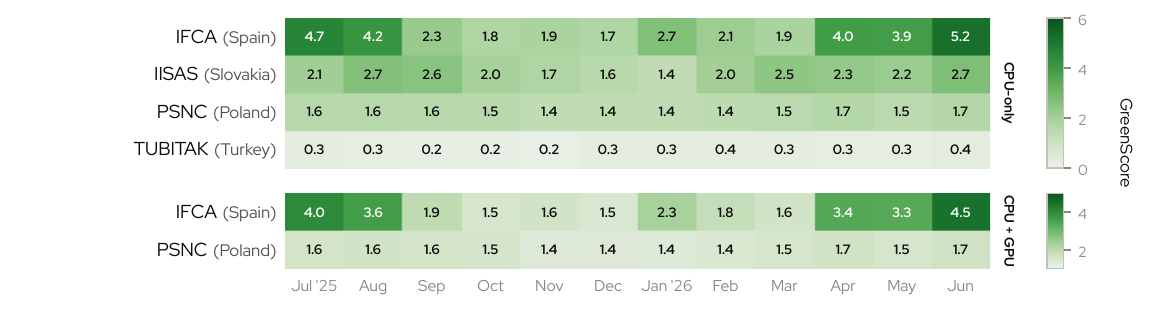}
    \caption{Monthly \gls{gs} for the AI4EOSC providers, split by workload profile: CPU-only (top, all four providers, hardware term $CEE_{n}$) and CPU+GPU (bottom, only the \gls{gpu}-equipped providers, hardware term \gls{hee}). Each cell is the arithmetic mean over the month of $GS_i(t)=HEE_i/[PUE_i\,(1-\mathrm{ES}_i(t)/100)]$, computed from the \gls{es} returned by Wattnet. The two blocks are normalised and coloured independently (each on a linear scale), since their hardware-efficiency terms are ratio-scaled over different provider sets; more saturated greens indicate a higher \gls{gs}.}
    \label{fig:gs}
\end{figure*}

In AI4EOSC, the \gls{gs} is evaluated per eligible compute node rather than per provider: when a job is submitted, the \gls{wms} calls \gls{gd} to rank the subset of nodes the user can access (Section~\ref{sec:green-director}), and each node carries its own \gls{hee}, built from the CPU model it runs and, on \gls{gpu}-equipped nodes, its \gls{gpu} model. The workload profile filters this set before ranking: a CPU-only job considers every eligible node, whereas a CPU+GPU job considers only the \gls{gpu}-equipped ones. Figure~\ref{fig:gs} summarises this node-level metric at provider granularity, reporting, per workload profile, the average \gls{gs} over a provider's nodes under the equal-node-count-per-model assumption above.

The \acrfull{gs} is therefore defined as shown in Equation~\ref{eq:gs}:

\begin{equation}\label{eq:gs}
    GS_i(t) = \frac{HEE_i(t)}{PUE_i(t) \cdot \left(1-\mathrm{ES}_i(t)/100\right)}
\end{equation}

where $\mathrm{ES}_i(t)$ is the \gls{es} defined in Equation~\ref{eq:es} (Section~\ref{sec:score}), $1-\mathrm{ES}_i(t)/100$ is the proportional combined environmental burden underlying that score, and $PUE$ is the \gls{pue}~\cite{noauthor_isoiec_nodate}. Unlike \gls{es}, which is a bounded, zone-relative ranking intended for reporting and comparison purposes, $GS$ is built directly from this unbounded, proportional quantity so that it remains proportional to the real physical impact of the workload's execution. This proportionality is essential for scheduling: since scheduling decisions optimise the amount of useful computational work delivered per unit of real impact (e.g. per gram of \ce{CO2} or per stress-litre of water), the metric driving those decisions must scale linearly with the underlying physical quantity, not merely move in the same direction as it. $GS$ therefore represents the amount of useful computational work that can be delivered per unit of real environmental impact at a given \gls{dc}. A higher \gls{gs} means a better overall environmental efficiency, and ``greener'' execution \glspl{dc} can be prioritised in scheduling decisions.

\begin{figure*}
    \centering
    \includegraphics[width=\textwidth]{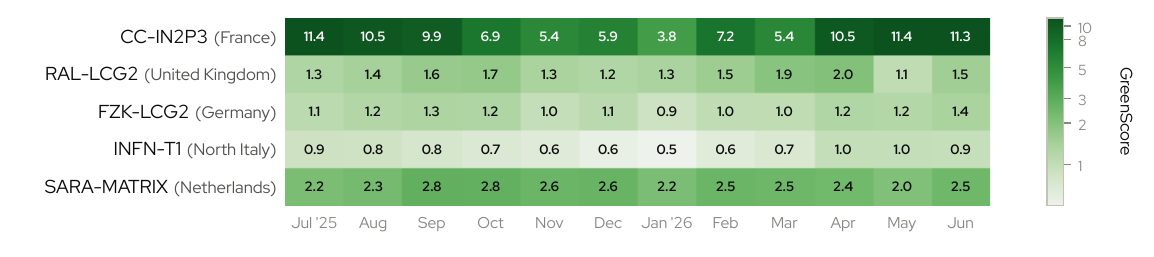}
    \caption{Monthly \gls{gs} for the DIRAC \gls{wms} sites, computed from the \gls{es}   as $GS_s(t)=CEE_{s}/[PUE_s\,(1-\mathrm{ES}_s(t)/100)]$, where $\mathrm{CEE}_{s}$ is the CEE of site $s$, normalized by the
maximum CEE observed across the five evaluated DIRAC sites; each cell is the arithmetic mean of $GS_s(t)$ over the month. The colour scale is logarithmic because CC-IN2P3, in France's low-carbon zone, stays well above the other sites. Higher values are preferable.}
    \label{fig:monthly-gs}
\end{figure*}

Figures~\ref{fig:gs} and~\ref{fig:monthly-gs} show the monthly \gls{gs} for the two use cases, computed from the \gls{es} that Wattnet publishes every 15 minutes for each provider's electricity zone and aggregated to a monthly arithmetic mean of $GS(t)$. They are shown separately because the two use cases draw on disjoint sets of candidate sites, and the hardware-efficiency term is ratio-scaled within each of them (over the AI4EOSC providers eligible for the workload profile, or over the five DIRAC \gls{wms} sites); cells are therefore comparable within a use case, and within a row group, but not across them.

Figure~\ref{fig:gs} groups the AI4EOSC providers by workload profile. For CPU-only workloads (top block), all four providers are candidates: IFCA leads, helped by a clean and increasingly solar-driven Spanish grid, and shows a marked summer peak; IISAS follows; PSNC stays flat despite efficient AMD EPYC-Genoa CPUs because of its carbon-intensive grid; and TUBITAK is far behind, penalised by both an old CPU generation and a carbon-intensive mix. For CPU+GPU workloads (bottom block), only IFCA and PSNC are eligible: PSNC has the higher hardware term (H100 \glspl{gpu}, lowest \gls{pue}), but its grid still keeps its \gls{gs} below IFCA's for most of the year. Each block is normalised and coloured independently, so cells compare within a block, not across the two.

Figure~\ref{fig:monthly-gs} shows the monthly \gls{gs} values calculated for five representative DIRAC sites: CC-IN2P3 (France), RAL-LCG2 (United Kingdom), FZK-LCG2 (Germany), INFN-T1 (North Italy), and SARA-MATRIX (Netherlands). CC-IN2P3 consistently obtains the highest score, although it also exhibits the largest monthly variation. SARA-MATRIX remains comparatively stable and occupies the second position, followed by RAL-LCG2, FZK-LCG2, and INFN-T1. The ordering remains unchanged throughout the observation period, despite temporal variations in the monthly scores. These differences reflect the combined effect of each site's \gls{cee} (ratio-scaled over the five \gls{wms} sites), \gls{pue}, and time-varying environmental conditions represented by \gls{es}. Higher values indicate a more favourable expected ratio of normalized CPU work to combined environmental burden.

\section{Practical implementation}
\label{sec:implementation}

\subsection{AI4EOSC allocation placement}
\label{sec:nomad-alloc}

The AI4EOSC platform \cite{HEREDIA2026108672} is a federated system that provides a cloud-based toolbox to build, train, share, and serve \gls{ai} models, exploiting distributed resources from pan-European e-Infrastructures. In the AI4EOSC platform, the \gls{wms} manages the execution of user tasks across different service providers. The \gls{wms} leverages different open-source components, namely Consul to federate the providers and to manage the platform state, and Nomad to run the workloads. The federation is organized in ``data centers'' (where one data center represents one cloud or resource provider) that provide resources to one or more namespaces. Each data center is composed of one or more nodes (Nomad clients) that are the ones executing the workloads. The namespaces are a logical organization of resources, allowing to segment different resources and users within a data center. In this regard, the tenancy model is based on \glspl{vo} (representing a group of users), and therefore the federated resources in a given data center can be dedicated to a single \gls{vo} (i.e., assigned to a single namespace) or shared among them. The current federation includes resources from four different scientific cloud providers located in four different countries: IFCA (Spain), IISAS (Slovakia), TUBITAK (Turkey) and PSNC (Poland).

The allocation placement in the AI4EOSC \gls{wms} is split into two distinct phases: feasibility checking and ranking.
In the first phase, the scheduler finds nodes that are feasible by enforcing hard constraints. There are two types of such constraints:

\begin{itemize}
    \item \textit{Internal constraints}: this means filtering nodes in data centers and node pools not used by the job, unhealthy nodes, and those missing necessary drivers.
    Since AI4EOSC operates with a single pool of nodes (though organized by namespaces) and all nodes have equivalent configurations in terms of drivers, this internal set of requirements typically enforces that the final allocation node is healthy and has enough free resources to accommodate the job specifications in terms of CPU/\gls{gpu}/RAM/Disk, including the usage of specific \gls{gpu} models.

    \item \textit{AI4EOSC platform constraints}: those are additional hard requirements that AI4EOSC enforces on each job.
    AI4EOSC implements the following constraints on the user jobs:

    \begin{itemize}
        \item Jobs can only be executed in client nodes that have passed the AI4EOSC integration tests (node status is \texttt{ready}). This ensures that the nodes are not only healthy from the Nomad perspective, but are also correctly configured according to the AI4EOSC specification.
        \item Jobs must only land on nodes serving the namespace serving the \gls{vo} of the user (because not all nodes serve all \glspl{vo}).
        \item They must only be deployed on nodes reserved for that specific kind of deployment the job corresponds to. AI4EOSC deployment types include: \texttt{compute} (for interactive jobs or long-running tasks), \texttt{batch} (for batch executions), and \texttt{tryme} (for short-lived inference endpoints).
    \end{itemize}

\end{itemize}

The second phase is a ranking, where the scheduler scores feasible nodes to find the best fit. In order to do so, the \gls{wms} leverages the Nomad scoring algorithm, which is primarily based on bin packing, which is used to optimize the resource utilization and density of applications, but is also augmented with affinity and anti-affinity rules. Nomad automatically applies a job anti-affinity rule which discourages collocating multiple instances of a task group. The combination of this anti-affinity and bin packing optimizes for density while reducing the probability of correlated failures.
On top of this Nomad scoring, AI4EOSC enforces its own set of soft requirements: those are usage patterns that AI4EOSC wishes to promote, but that should not block from being deployed if they cannot be fulfilled. AI4EOSC affinities include the following:

\begin{itemize}
    \item If the job only requires CPU resources, it will try to avoid \gls{gpu} nodes (to avoid filling \gls{gpu} nodes with jobs that do not require \glspl{gpu}).
    \item It will try to land on nodes that exclusively serve that namespace (instead of landing in wildcard nodes that can serve different \glspl{vo} at the same time).
    \item It will try to land on greener data centers. This affinity is computed by the \textit{\gls{gd}}, described in Section~\ref{sec:green-director}.
\end{itemize}

\subsubsection{AI4EOSC \acrlong{gd}}
\label{sec:green-director}

AI4EOSC computes green affinities using the \acrfull{gd} component, inside the AI4EOSC Platform \gls{api}. The typical workflow operates as follows.
First, \gls{gd} calls the Wattnet \gls{api} \footnote{\url{https://api.wattnet.eu}} to retrieve the \gls{es} for each AI4EOSC data center, based on the data center coordinates. If a value cannot be retrieved (e.g., because the location is outside Europe), a reasonable default value is used instead. This value is kept as part of the data center state; therefore, when a new user job is about to be launched, the AI4EOSC \gls{wms} calls the \gls{gd} to rank the subset of data center nodes the user has access to (as already mentioned, the nodes visible to the user depend on the user \gls{vo}). For each data center, \gls{gd} averages the \gls{es} over the last 7 days to have a smooth short-term value. Finally, the averaged \gls{es}, the data center \gls{pue}, and each node's \gls{hee} (Equation~\ref{HEE}, reduced to its \gls{cee} term for CPU-only workloads) are combined according to Equation~\ref{eq:gs} to assign a per-node \gls{gs}.

Using those scores $v$, the \gls{gd} then ranks the nodes on a $[0, 100]$ affinity scale using a linear fit, where the maximum score $v_{\max}$ maps to the maximum affinity $af_{\max} = 100$ and the minimum score $v_{\min}$ maps to the minimum affinity $af_{\min} = 0$:

\begin{equation}
    DC_{\text{affinity}} = af_{\min} + (af_{\max} - af_{\min}) \times \frac{(v - v_{\min})}{(v_{\max} - v_{\min})}
\end{equation}

While a linear interpolation is used for simplicity, the \gls{gd} can easily be extended to use other ranking algorithms.
Finally, the $[0, 100]$ ranking is rescaled to a $[0, 30]$ range in order to soften the green affinity requirement with respect to the other two AI4EOSC affinities (\ref{sec:nomad-alloc}). Based on the new scaled values, one affinity per node is created in the Nomad job.

\subsubsection{Validation}

Validating the \gls{gd} in a real production setup is not straightforward: as seen in Section~\ref{sec:nomad-alloc}, the job placement depends on multiple dynamical factors, in particular the current load of the different nodes, which varies constantly.

We propose two validation strategies:

\begin{itemize}

    \item \textit{Score inspection}: In this approach, we launch the same version of the job with different values of a green affinity. We then inspect the total ranking score (combining binpack and affinity) that the \gls{wms} gave to each node for that job. After each try, we delete the job to avoid impacting the load distribution (otherwise the binpack ranking score of the subsequent jobs would be modified, invalidating the comparison). For the same reason, we run this experiment in a short time span to avoid other user jobs being allocated while the experiment is running, thus modifying the binpack score of a given node.   
    To minimize the noise caused by the resource availability constraint, we use a job with very little resource requirements, thus making sure it will fit in all available nodes. 
    
    \item \textit{Cluster filling}: In this approach, we ask ourselves: if the cluster were suddenly flooded with new jobs, would the green affinity indeed make the greener data centers fill up first? To put this in practice, we send in bulk as many jobs as needed to fill the whole cluster and record the usage of the different data centers. To minimize impact on production users, we run the experiment in a low-demand period (weekend night) and immediately delete the deployments afterwards. To maximize the effect of the green affinity, we set it to the maximum value ($100$). All launched jobs have the same hardware requirements.

\end{itemize}

For both validation strategies, to see more clearly the effects of the green affinity, we only add the green affinity in a single data center (\texttt{tubitak-imagine}), setting to zero the green affinities of other data centers.

\subsection{DIRAC Pilot (Worker) allocation placement}
\label{sec:DIRAC}
\begin{figure}[H]
    \centering
    \includegraphics[width=0.98\columnwidth]
    {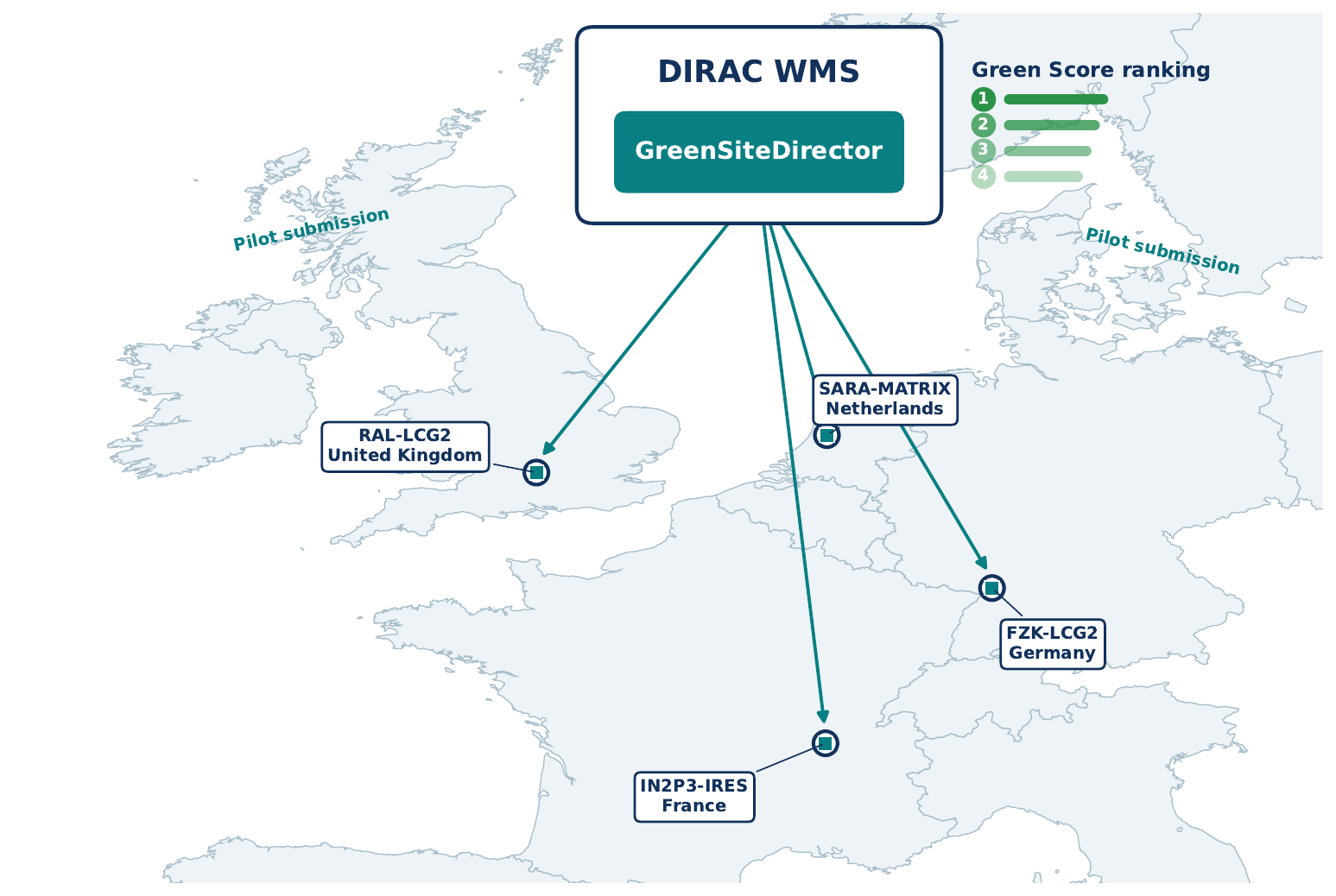}
    \caption{\acrfull{gsd}-based pilot (worker) placement across the
    evaluated DIRAC sites. Eligible data centers are ranked by \acrfull{gs} before pilot submission.}
    \label{fig:dirac-greensitedirector}
\end{figure}

The DIRAC Interware supports High-Throughput Computing workflows
composed of large numbers of independent or loosely coupled jobs
executed across federated data centers or computing sites as shown in Figure \ref{fig:dirac-greensitedirector}. DIRAC provides workload, data, and production-management
services across heterogeneous distributed computing
resources~\cite{stagni_dirac_2020,stagni_dirac_2024,
stagni_diracx_2025}. Its source code is developed openly under
the GPLv3 licence~\cite{diracgrid_github}.

DIRAC follows a
pilot-based execution model: the \textit{SiteDirector} submits pilot jobs (worker slots) to different \glspl{dc}, and running pilots retrieve user payloads (jobs) from
the central queue. Within this pilot-based model, environmental prioritization is introduced by modifying the order in which the \textit{SiteDirector} considers eligible computing sites.

\subsubsection{DIRAC \acrfull{gsd}}
\label{sec:dirac-gd}
In the baseline DIRAC provisioning strategy, pilot submission is
driven primarily by workload demand. The \textit{SiteDirector} monitors the
number of waiting jobs and submits pilots to eligible data centers to
satisfy that demand. Candidate data centers are considered in a
non-deterministic, typically randomized, order during each scheduling
cycle.

For each data center, the number of pilots that can be submitted
depends on the remaining workload demand, the number of pilots already
running or starting, the site-specific pilot capacity, and the
compatibility between the workload and the site's capabilities.
Pilots are submitted greedily until either the demand is satisfied or
no additional compatible capacity is available.

This strategy promotes throughput, resource utilization, and load balancing across data centers, but it
does not differentiate data centers according to their environmental
characteristics. In particular, the baseline ordering does not
consider sites'  energy efficiency,  carbon intensity,
water-scarcity impact, or \gls{pue}. Consequently, the same
computational workload may be executed at sites with substantially
different environmental impacts.

To address this limitation, we developed a customized \textit{SiteDirector},
referred to as the \acrfull{gsd}. It preserves DIRAC's
demand-driven pilot-provisioning mechanism but introduces an explicit
environmental prioritization of candidate data centers. The central
difference from the baseline strategy is the order in which sites are
considered before pilot submission.

For each eligible data center, the \acrfull{gsd} obtains
the \acrfull{gs} defined in Equation~\ref{eq:gs}. For the CPU-only DIRAC
workloads considered here, the hardware-efficiency term is represented
by the site's \gls{cee}, as defined in Equation~\ref{CEE}. The CEE values are normalized by the maximum CEE observed across the evaluated DIRAC sites. Accordingly, the site-specific \gls{gs} used by the
\acrshort{gsd} is calculated as
\begin{equation}
\label{eq:dirac-gs} 
GS_s(t)=
\frac{CEE_s(t)}
{PUE_s(t)\left(1-ES_s(t)/100\right)},
\end{equation}
where $ES_s(t)$ is supplied by Wattnet~\cite{castrillo_melguizo_wattnet_2026},
$CEE_s(t)$ represents the CPU energy efficiency of site $s$ at
time $t$, and $PUE_s(t)$ is its Power Usage Effectiveness.
A higher \gls{gs} indicates that the site is expected to deliver more
normalized CPU work per unit of combined environmental impact.
The scheduler uses this quantity directly; in Figure~\ref{fig:monthly-gs} the \gls{cee} term is additionally ratio-scaled over the five sites (Equation~\ref{scaling}) for readability, which, being a division by a constant, leaves the site ordering unchanged.

The \acrshort{gsd} scheduling cycle proceeds as follows:

\begin{itemize}[
    leftmargin=*,
    itemsep=2pt,
    topsep=4pt,
    parsep=1pt,
    partopsep=0pt
]
    \item The \gls{gs} is obtained for each eligible data center.
    \item Candidate data centers are sorted in decreasing \gls{gs} order.
    \item Pilots are submitted greedily following this ordering,
    considering:
    \begin{itemize}[nosep]
        \item the number of waiting jobs;
        \item the number of pilots already running or starting;
        \item the pilot capacity available at each data center; and
        \item compatibility between workload requirements and site capabilities.
    \end{itemize}
    \item Pilot submission stops when the workload demand is satisfied
    or no additional compatible pilot capacity is available.
\end{itemize}

The \gls{gs} is used as a scheduling preference rather than as a
hard placement constraint. Therefore, lower-ranked data centers may
still receive pilots when higher-ranked sites reach capacity or cannot
satisfy the workload requirements. This preserves DIRAC's fundamental
scheduling semantics, including demand-driven provisioning and workload compatibility, while preferentially directing
pilots towards data centers expected to deliver more normalized
computation per unit of combined environmental impact.

\subsubsection{Validation}
\label{sec:dirac-validation}
The  \acrfull{gsd} is evaluated in two complementary
settings. First, we perform a controlled trace-driven simulation
that replays a workload extracted from DIRAC job history. Each computing site is represented using empirical hardware and
performance characteristics derived from accounting records of jobs
executed at the site during the preceding 12 months. These include the
average CPU normalization factor, worker-node core count, and \gls{tdp}. The
site's \gls{pue} and historical \acrfull{gs}, calculated from its \gls{cee} and
Wattnet environmental data, complete this characterization. Both
scheduling policies use the same workload, site capacities, site
characteristics, and environmental data; only the site-ordering policy
changes. This provides a reproducible comparison between randomized
and \gls{gs}-based scheduling.

Second, we present a preliminary evaluation of the
\acrshort{gsd} in the production DIRAC infrastructure using
jobs from the KM3NeT  (a European deep-sea neutrino research
collaboration) \gls{vo}. This evaluation
complements the controlled simulation by showing the behaviour
of \gls{gs}-based pilot ordering under real operating
conditions, including changing workload demand, site
availability, local batch queues, pilot startup and failure rates,
and workload compatibility constraints. Because the production
observation periods are not experimentally controlled, these
results are interpreted as initial operational evidence rather
than as a causal policy comparison.

\paragraph{Trace-driven simulation.}
The simulator operates as a discrete-time loop with a one-minute
resolution. At every tick, it releases jobs according to their
submission times in the historical trace, assigns waiting jobs to
available pilot slots, advances running jobs, records completed jobs,
and updates the simulation clock.

Each configured site slot represents one persistent, implicit DIRAC
pilot (worker) that executes at most one job at a time. Pilot startup delay
is assumed to be zero, and all configured pilot slots are initially
available. When a job finishes, its pilot slot becomes available and
acquires the next job from the FIFO waiting queue during the following
simulation tick. If the queue is empty, the slot remains idle. Idle
pilot slots are represented in the capacity model but do not receive
a separate energy charge.


Each data center  is characterized by its average server TDP, number
of cores per server, and CPU normalization factor. These values are
derived from historical jobs executed at the site by averaging the
reported worker-node TDPs, core counts, and CPU normalization factors.
They represent the typical worker-node configuration historically
observed at each site and remain fixed under both scheduling policies.
Table~\ref{tab:dirac-notation} summarizes the notation used in the
following formulation.
\begin{table}[htbp]
\centering
\caption{Notation used in the DIRAC trace-driven simulation.}
\label{tab:dirac-notation}
\scriptsize
\renewcommand{\arraystretch}{1.12}
\begin{tabular}{p{0.19\columnwidth}p{0.73\columnwidth}}
\toprule
Symbol & Description \\
\midrule

$j$, $s$ &
Job and selected computing site, respectively. \\

$N_j$ &
Normalized CPU demand of job $j$, expressed in normalized
CPU-seconds. \\

$W_j$ &
Historical wall-clock duration of job $j$, expressed in seconds. \\

$q_j$ &
CPU normalization factor reported for the job's original worker
node. \\

$\phi_s$ &
Average CPU normalization factor of site $s$; higher values indicate
faster execution. \\

$C_{j,s}$ &
Estimated CPU time of job $j$ at site $s$, expressed in seconds. \\

$W_{j,s}$ &
Estimated wall-clock duration of job $j$ at site $s$, expressed in
seconds. \\

$R_{j,s}$ &
Number of one-minute simulation intervals for which job $j$ occupies
a pilot slot at site $s$, expressed in minutes. \\

$n_s$ &
Average number of processor cores per worker node at site $s$. \\

$P_s$ &
Average worker-node \gls{tdp} at site $s$, expressed in watts. \\

$f$ &
Dimensionless non-active power factor during job execution;
$f=0.4$. \\

$E_{j,s}$ &
Energy attributed to job $j$ at site $s$ before facility overhead,
expressed in kilowatt-hours. \\

$\mathrm{PUE}_s$ &
Power Usage Effectiveness of site $s$, treated as a constant
dimensionless factor during the simulation. \\

$\mathrm{CI}_s(t)$ &
Average carbon intensity at site $s$ over the historical execution
interval reconstructed for job $j$ from the trace, expressed in grams
of CO$_2$e per kilowatt-hour. \\

$\mathrm{WI}_s(t)$ &
Average AWARE-weighted water-impact intensity at site $s$ over the
historical execution interval reconstructed for job $j$ from the
trace, expressed in stress-L per kilowatt-hour. \\

$CF_{j,s}$ &
Carbon footprint of job $j$ at site $s$, expressed in kilograms of
CO$_2$e. \\

$HI_{j,s}$ &
AWARE-weighted water-scarcity impact of job $j$ at site $s$,
expressed in stress-L. \\

$\mathrm{GS}_s$ &
Historical GreenScore used to rank site $s$; higher values are
preferable. \\

\bottomrule
\end{tabular}
\end{table}

When job $j$ is assigned to data center $s$, its CPU time and wall-clock
duration are adjusted according to the performance of the selected
site:

\begin{align}
C_{j,s} &= \frac{N_j}{\phi_s},\\
W_{j,s} &= \frac{W_j q_j}{\phi_s},\\
R_{j,s} &= \max\left(
1,
\left\lceil\frac{W_{j,s}}{60}\right\rceil
\right),
\label{eq:dirac-runtime}
\end{align}

where $N_j$ is the normalized CPU demand of the job, $W_j$ is its
historical wall-clock duration, $q_j$ is the CPU normalization factor
reported for its original worker node, and $\phi_s$ is the average
CPU normalization factor of the selected site. The resulting value
$R_{j,s}$ determines how many one-minute simulation ticks the pilot
slot remains occupied.

Because the workload is single-core, the energy attributed to job
$j$ at site $s$ is estimated as

\begin{equation}
E_{j,s} =
\frac{\left[(1-f)C_{j,s}+fW_{j,s}\right]}{3.6\times10^6}
\frac{P_s}{n_s},
\qquad f=0.4,
\label{eq:dirac-energy}
\end{equation}

where $P_s$ is the average server TDP at site $s$, $n_s$ is its
average number of cores per server, and $f$ is the assumed fraction
of server power associated with the non-active power component. We
set $f=0.4$ for all sites. The value $E_{j,s}$ represents the
job-attributed energy consumption before accounting for data-center overhead.

The carbon footprint and AWARE-weighted water-scarcity impact are
calculated as

\begin{align}
CF_{j,s} &=
E_{j,s}\,\mathrm{PUE}_s\,
\frac{\mathrm{CI}_s(t)}{1000},
\label{eq:dirac-carbon}\\
HI_{j,s} &=
E_{j,s}\,\mathrm{PUE}_s\,
\mathrm{WI}_s(t),
\label{eq:dirac-water}
\end{align}

where $\mathrm{CI}_s(t)$ and $\mathrm{WI}_s(t)$ are averaged over the
historical execution interval associated with job $j$ in the original
trace. The factor $\mathrm{PUE}_s$ accounts for facility-level energy
overhead at the selected data center. Since carbon intensity is
expressed in gCO$_2$e/kWh, division by 1000 gives $CF_{j,s}$ in
kgCO$_2$e. Similarly, $HI_{j,s}$ is expressed in stress-L.

Two scheduling policies are compared. The baseline policy randomly
shuffles the candidate data centers during each scheduling cycle. The \gls{gs}-based policy instead sorts them in descending order of
their fixed historical  \acrfull{gs}, before filling the available
pilot slots. In both cases, jobs remain in FIFO order, and the
number of jobs assigned to each data center is limited by its available
pilot capacity and the remaining waiting demand.

Pilot allocation continues until either all waiting jobs have been
assigned or every eligible data center has reached its pilot capacity.
The simulation terminates once every job in the workload trace has
completed.

\paragraph{Preliminary production evaluation}
Following the controlled simulation, we performed a preliminary
evaluation of the \acrfull{gsd} in the DIRAC production infrastructure.
Whereas the simulation replays the same workload and operating
conditions for each policy, the production environment is subject to
uncontrolled variations in workload demand, site availability,
pilot-startup delays and failure rates, local batch queues (local site
schedulers), and workload-compatibility constraints. The results
should therefore be interpreted as initial operational evidence rather
than as a fully controlled comparison between scheduling policies.

The preliminary production results, together with the trace-driven
simulation results, are presented and discussed in
Section~\ref{sec:validation-results-dirac}.

\section{Results and discussion}
\label{sec:results}

\subsection{Environmental characterization of electricity consumption by the infrastructure providers}
\label{sec:env-characterization}

\begin{figure}[htbp]
    \centering
    \includegraphics[width=\linewidth]{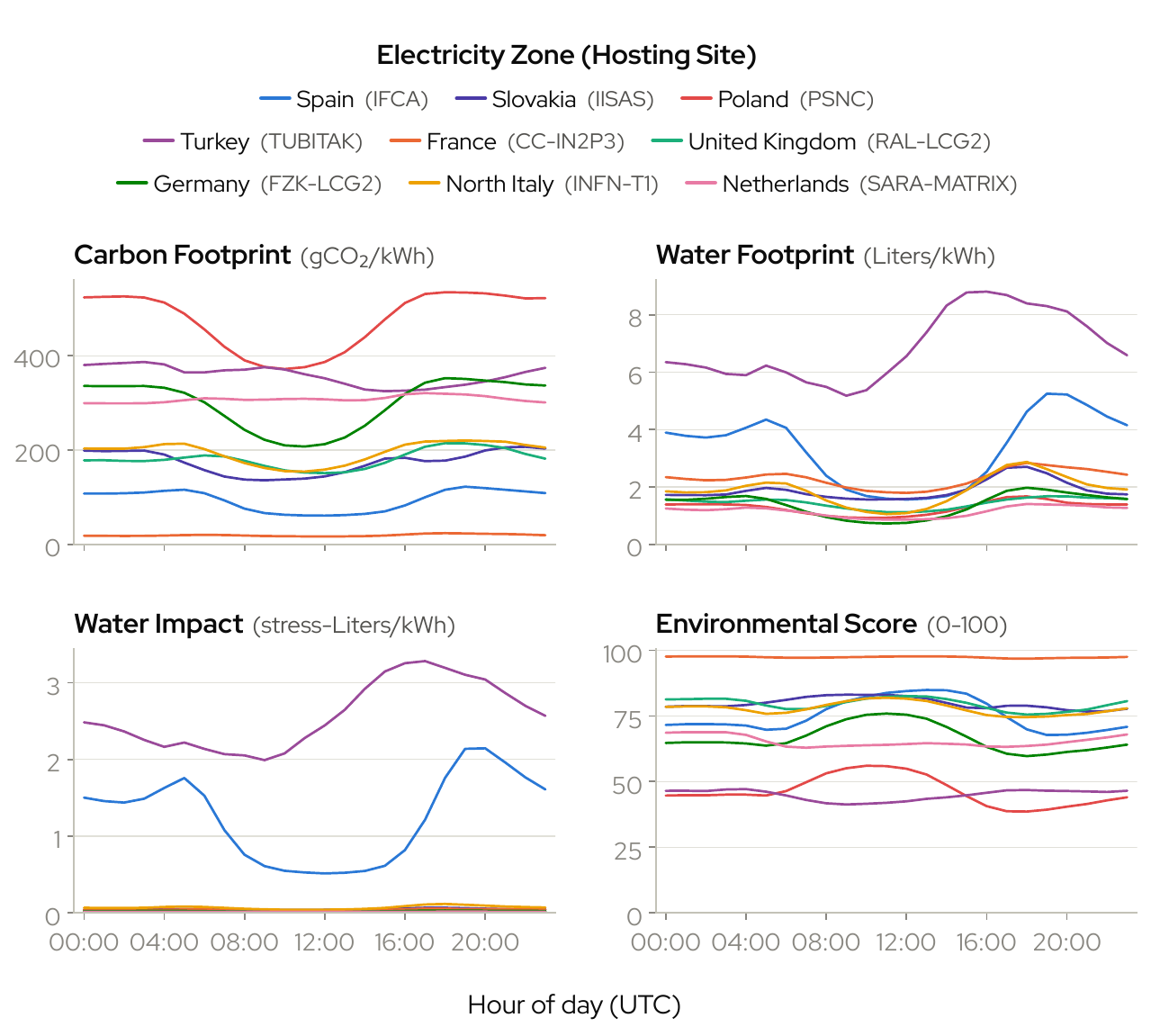}
    \caption{Average diurnal profile (UTC) of \gls{cf}, \gls{wf},
    \gls{wi} and \gls{es} for the nine infrastructure providers of the two
    use cases: the four AI4EOSC providers (IFCA/Spain, IISAS/Slovakia,
    PSNC/Poland, TUBITAK/Turkey) and the selected DIRAC \gls{wms} sites, one
    per country (CC-IN2P3/France, RAL-LCG2/United Kingdom, FZK-LCG2/Germany,
    INFN-T1/North Italy, SARA-MATRIX/Netherlands), aggregated over the
    period from July 2025 to June 2026 at 15-minute resolution.}
    \label{fig:hourly_profile}
\end{figure}

\begin{figure}[htbp]
    \centering
    \begin{subfigure}[b]{0.48\linewidth}
        \includegraphics[width=\textwidth]{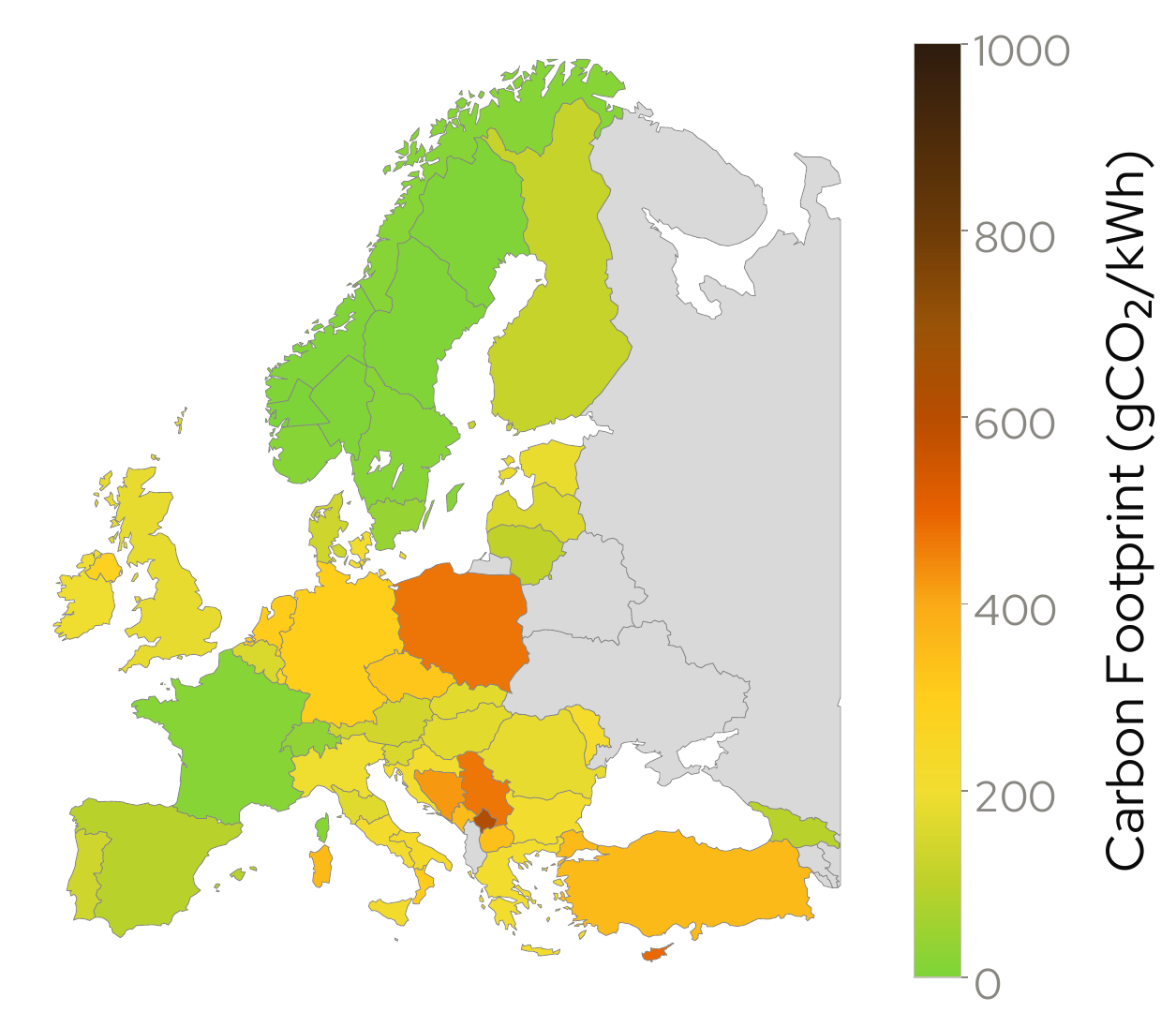}
        \caption{\gls{cf} (gCO$_2$/kWh)}
        \label{fig:map_cf}
    \end{subfigure}
    \hfill
    \begin{subfigure}[b]{0.48\linewidth}
        \includegraphics[width=\textwidth]{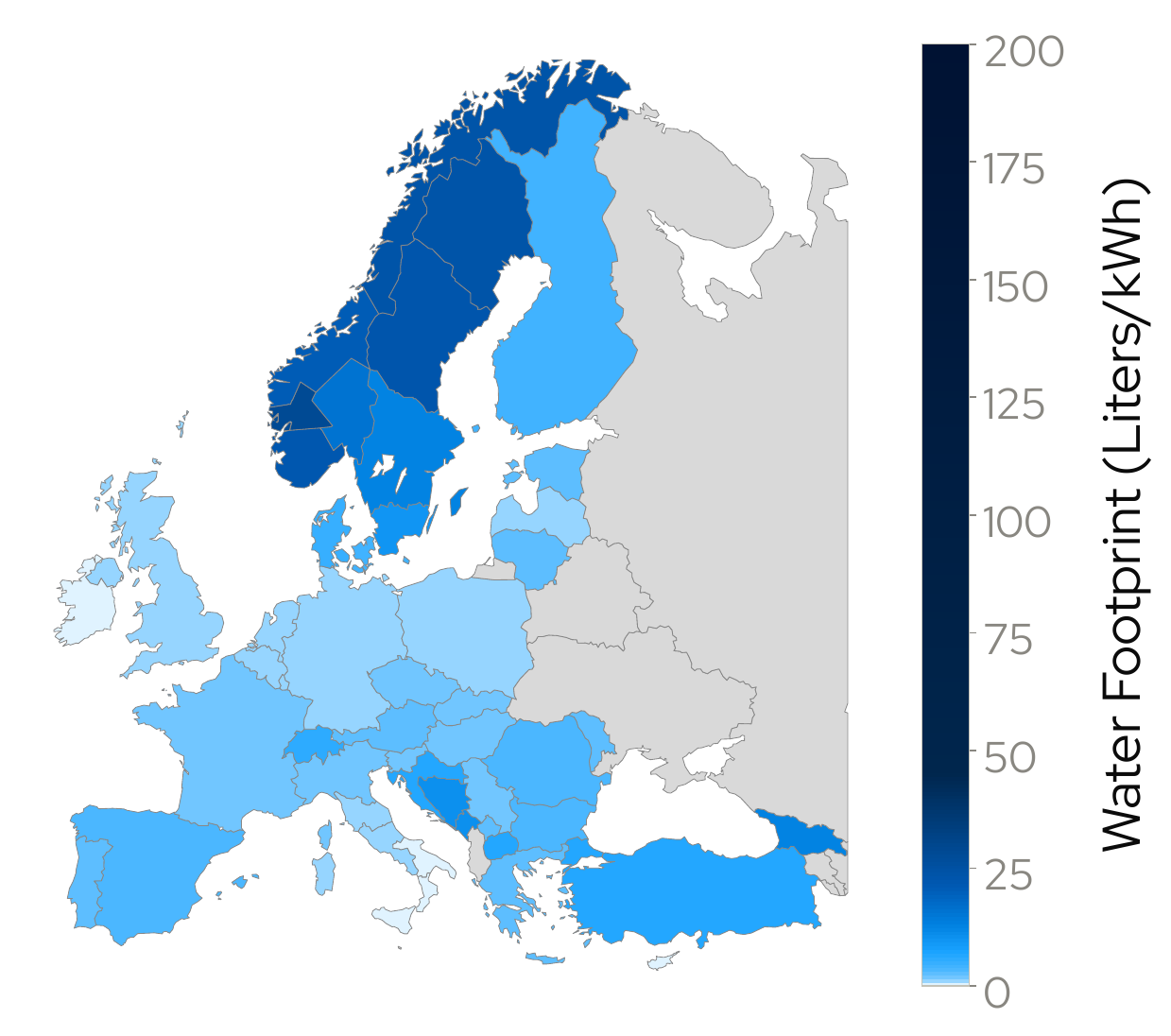}
        \caption{\gls{wf} (L/kWh)}
        \label{fig:map_wf}
    \end{subfigure}
    \\[1em]
    \begin{subfigure}[b]{0.48\linewidth}
        \includegraphics[width=\textwidth]{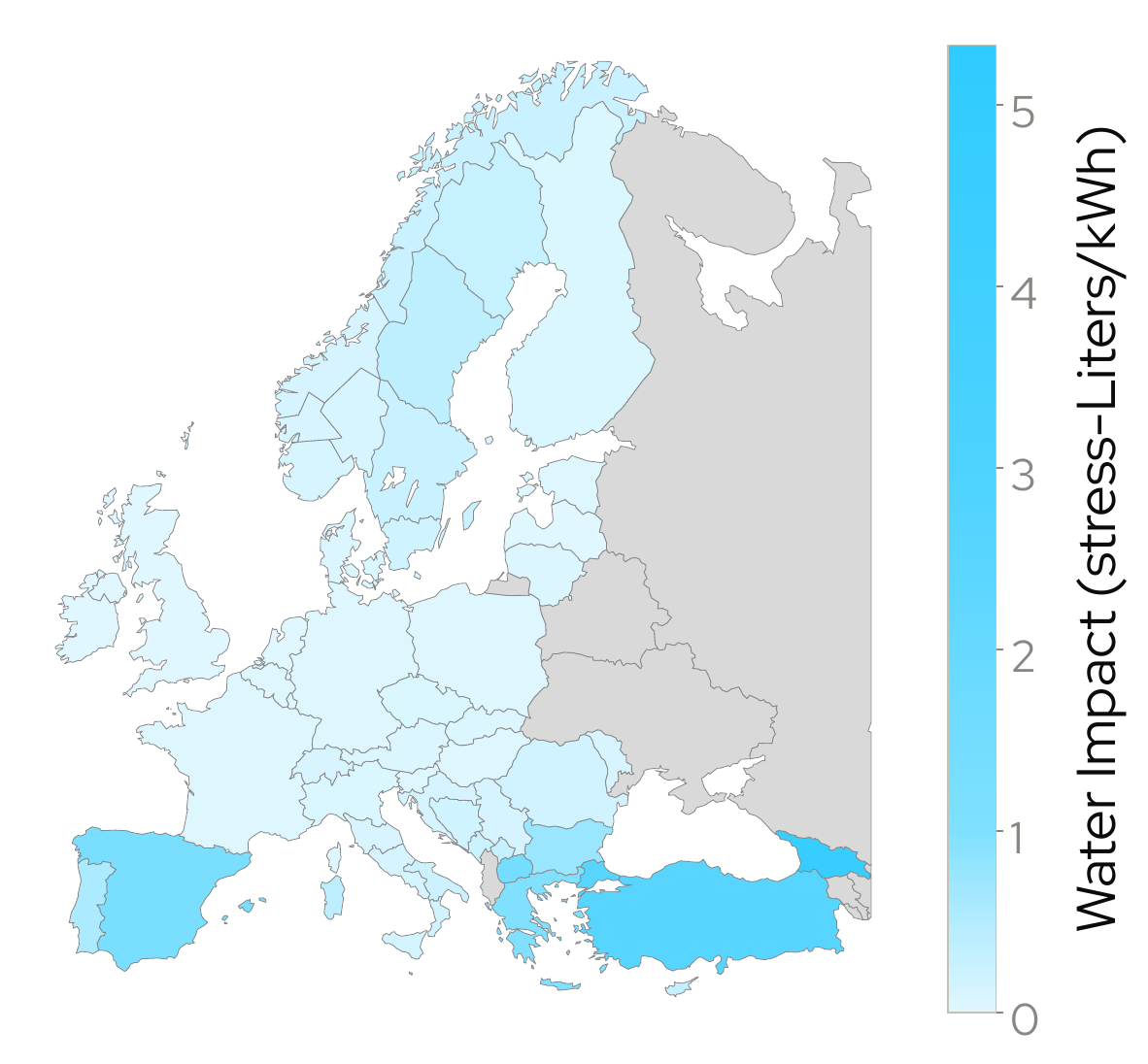}
        \caption{\gls{wi} (stress-L/kWh)}
        \label{fig:map_wi}
    \end{subfigure}
    \hfill
    \begin{subfigure}[b]{0.48\linewidth}
        \includegraphics[width=\textwidth]{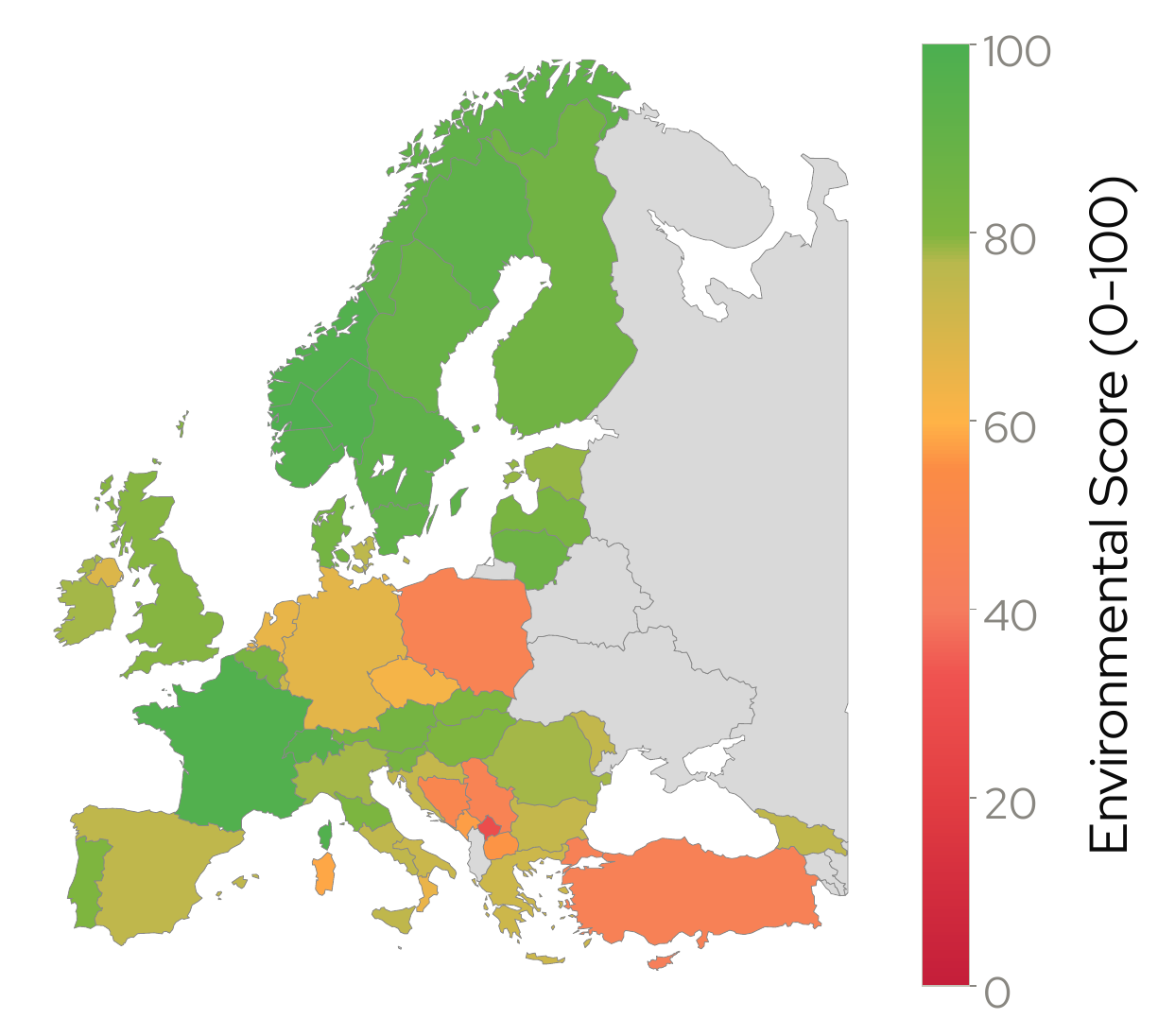}
        \caption{\gls{es} ($[0, 100]$)}
        \label{fig:map_score}
    \end{subfigure}
    \caption{Annual average (from July 2025 to June 2026) geographical distribution
    of the four environmental metrics across European bidding zones. Maps 
    (a)(b) show the raw carbon and \gls{wf}s of electricity 
    generation, while map (c) shows the \gls{wf} re-weighted by 
    the \gls{aware}2.0 scarcity factor. Map (d) shows the resulting unified 
    \gls{es}. Note the geographic mismatch between (b) and (c): countries 
    with high hydropower-driven \glspl{wf} (e.g., the Nordic region) 
    do not necessarily suffer high water stress, while Mediterranean zones 
    with moderate raw \glspl{wf} show severe stress-weighted impact. Among the
    nine providers of the two use cases, only IFCA and TUBITAK lie in a zone
    with appreciable \gls{aware}2.0-weighted water stress.}
    \label{fig:maps_grid}
\end{figure}

\begin{figure*}[htbp]
    \centering
    \includegraphics[width=\textwidth]{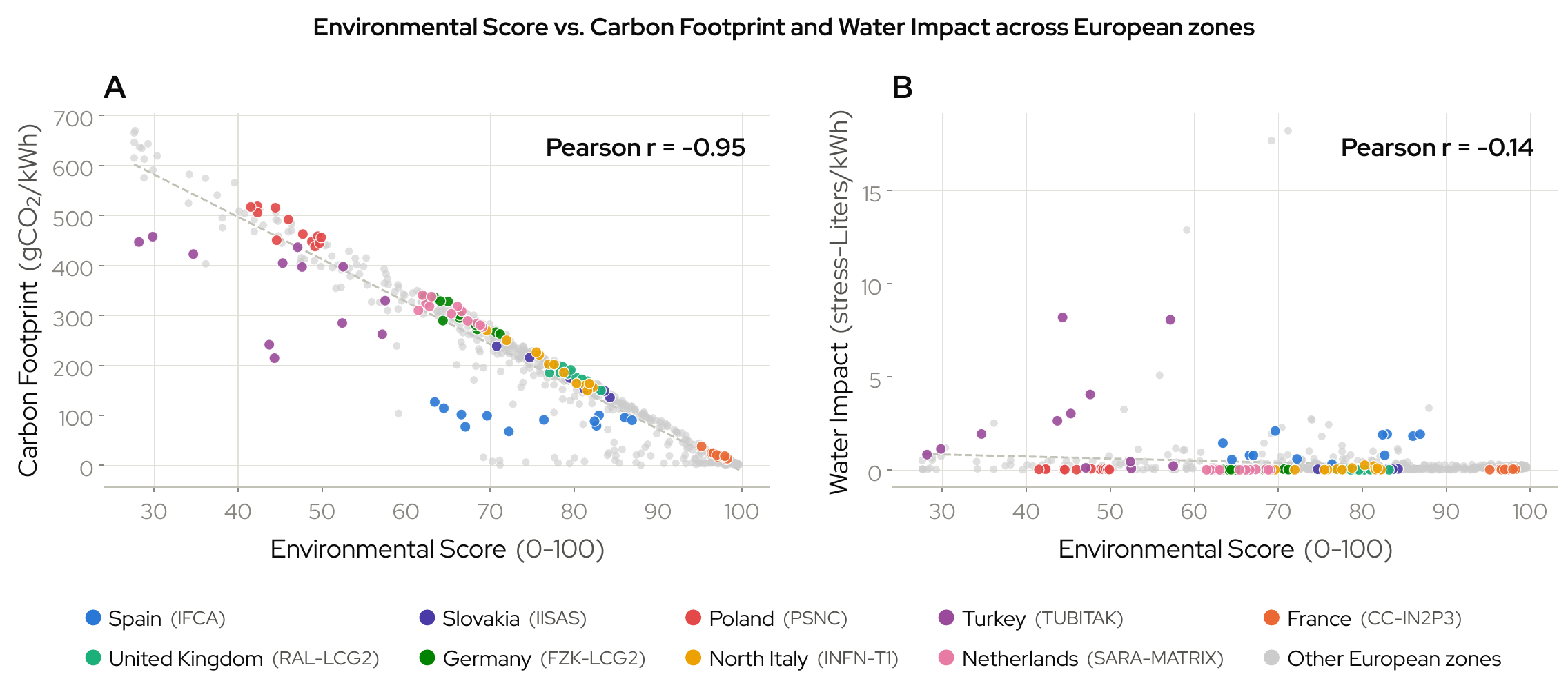}
    \caption{\gls{es} against \gls{cf} (panel A, Pearson $r=-0.95$) and
    against \gls{wi} (panel B, Pearson $r=-0.14$), with one point per
    European zone and month. Coloured markers are the nine AI4EOSC and DIRAC
    \gls{wms} providers; grey markers are the remaining European zones. The
    dashed line in each panel is the linear fit over all zones.}
    \label{fig:scatter_pair}
\end{figure*}

Before evaluating the ranking behaviour of the \gls{gd}, it is useful to characterise how environmentally heterogeneous the infrastructure providers of both use cases actually are, across time and across space. This heterogeneity is precisely what the green affinity described in Section~\ref{sec:green-director} is designed to exploit. We analyse jointly the nine providers introduced in Section~\ref{sec:es}, namely the four AI4EOSC platform providers (IFCA/Spain, IISAS/Slovakia, PSNC/Poland, TUBITAK/Turkey) and the five DIRAC \gls{wms} sites, one per country (CC-IN2P3/France, RAL-LCG2/United Kingdom, FZK-LCG2/Germany, INFN-T1/North Italy, SARA-MATRIX/Netherlands), along three complementary axes: the intra-day temporal variability of each metric (Figure~\ref{fig:hourly_profile}), its spatial distribution across European zones (Figure~\ref{fig:maps_grid}), and the relationship between the unified \gls{es} and the underlying \gls{cf} and \gls{wi} signals (Figure~\ref{fig:scatter_pair}). Throughout this section, the four metrics (\gls{cf}, \gls{wf}, \gls{wi} and \gls{es}) always refer to the electricity consumed by each of the nine providers, that is, to the operational, Scope-2-like footprint of the power drawn from the grid at each site, and not to any on-site facility resource use such as cooling water (Section~\ref{sec:score}). All values are computed from Wattnet operational data at 15-minute resolution over the period from July 2025 to June 2026, and account for cross-border electricity exchange (imports and exports) between the generating zone and the zone hosting each provider.

Figure~\ref{fig:hourly_profile} shows the average diurnal profile (in UTC) of the four metrics for the nine providers. \gls{cf} follows a common daily shape in most zones: a minimum around midday (from 10:00 to 13:00~UTC), when solar generation displaces fossil plants, and a maximum during the evening ramp (from 17:00 to 19:00~UTC), when solar output fades while demand stays high. The amplitude of this swing is, however, strongly zone-dependent. It is largest in absolute terms for the zones with the highest solar or coal share: PSNC and FZK-LCG2 vary by 163 and 145~gCO$_2$/kWh between their daily minimum and maximum, and IFCA nearly doubles its \gls{cf} (from 61 to 122~gCO$_2$/kWh, a swing of 65\% of its daily mean). By contrast, CC-IN2P3 is almost flat (17 to 24~gCO$_2$/kWh) because the French nuclear baseload has no daily solar cycle, and TUBITAK stays within a narrow band with a slight nocturnal maximum. \gls{wi} exhibits the same diurnal shape but with even larger relative amplitude where hydropower is part of the mix (a swing of 131\% of the daily mean for IFCA and 109\% for INFN-T1), because the midday solar surplus displaces precisely the water-intensive hydroelectric generation. For the DIRAC sites located in non-stressed zones, the diurnal structure of \gls{wi} is irrelevant in absolute terms (below 0.05~stress-L/kWh throughout the day). As a result, the \gls{es} is a near-mirror image of the \gls{cf} profile: it peaks around midday (up to 85 for IFCA at 13:00~UTC and 56 for PSNC at 10:00~UTC) and drops during the evening (down to 68 and 39, respectively), a daily swing of up to 17 \gls{es} points, whereas CC-IN2P3 remains at roughly 97 all day.

This temporal structure has a direct scheduling implication. For zones with a pronounced solar cycle (IFCA, PSNC, FZK-LCG2), shifting a delay-tolerant job into the window from 10:00 to 14:00~UTC improves its \gls{es} by up to 17 points without moving it to a different site; this is the margin that the temporal component of a carbon- and water-aware scheduler can exploit. For baseload-dominated zones such as CC-IN2P3, time-shifting yields essentially nothing and only spatial placement matters.

Figure~\ref{fig:maps_grid} shows the annual-average geographical distribution of the four metrics across the European zones covered by Wattnet. The \gls{cf} map (a) displays a clear gradient: the Nordic zones (below 6~gCO$_2$/kWh) and France (19~gCO$_2$/kWh) are almost carbon-free, whereas the coal-dependent zones of Central and South-Eastern Europe (Poland at 475, Cyprus at 489, Kosovo above 600~gCO$_2$/kWh) are an order of magnitude higher. The nine providers span most of this range: CC-IN2P3 sits at the clean end, IFCA, IISAS, RAL-LCG2 and INFN-T1 in the low-to-intermediate band, FZK-LCG2 and SARA-MATRIX in the upper-intermediate band, and PSNC and TUBITAK among the most carbon-intensive.

The comparison between the raw \gls{wf} map (b) and the scarcity-weighted \gls{wi} map (c) is the central observation of this characterization. Raw \gls{wf} is highest in the hydropower-rich Nordic zones (up to 29~L/kWh, an order of magnitude above the coal zones), simply because hydroelectric generation is water-intensive. Once the \gls{aware}2.0 scarcity factor is applied, this pattern almost completely inverts: the Nordic zones collapse to near zero, since their water is abundant, while the Mediterranean and semi-arid zones (Georgia at 4.6, Turkey at 2.6, North Macedonia at 1.4, Spain at 1.2~stress-L/kWh) come to dominate. Of the nine providers, only IFCA and TUBITAK lie in a zone with appreciable water stress; INFN-T1 is marginally elevated (0.07~stress-L/kWh), and the remaining six are all below 0.05~stress-L/kWh.

The \gls{es} map (d) is the synthesis of the two. Its large-scale structure follows the carbon gradient (the Nordic zones and France score highest), but the water correction is visible where it matters. A clear example among our providers is that IISAS obtains a higher \gls{es} than IFCA (80 versus 75) despite having almost twice its carbon intensity (175 versus 94~gCO$_2$/kWh), because IFCA carries a water-scarcity penalty (1.25 versus 0.05~stress-L/kWh) that IISAS does not. The \gls{es} is therefore not a simple re-ranking of the \gls{cf}.

Figure~\ref{fig:scatter_pair} plots the \gls{es} against the \gls{cf} (panel A) and against the \gls{wi} (panel B) for every European zone, with one point per zone and month and the nine providers highlighted. The two Pearson correlation coefficients differ sharply in magnitude ($r=-0.95$ for \gls{cf} and $r=-0.14$ for \gls{wi}), and this contrast is itself informative.

The strong correlation between \gls{es} and \gls{cf} is expected by construction. In Equation~\ref{eq:es}, the carbon term carries the larger weight ($W_C=0.71$), and across the European zone set the ratio-scaled carbon indicator $CF^{r}$ is broadly distributed (mean 0.32, standard deviation 0.24), because carbon intensity itself spans more than two orders of magnitude (from about 1 to about 620~gCO$_2$/kWh). The ratio-scaled water indicator $WI^{r}$, in contrast, is concentrated near zero: 58\% of all zone-months fall below $WI^{r}=0.05$ and 75\% below 0.10 (mean 0.10, standard deviation 0.19), with only a short tail of Mediterranean zones taking larger values. Decomposing the variance of the combined burden term $W_C\,CF^{r}+W_W\,WI^{r}$ accordingly, the carbon contribution accounts for roughly 90\% of it ($\operatorname{var}(W_C\,CF^{r})\approx0.028$ versus $\operatorname{var}(W_W\,WI^{r})\approx0.003$). To first order, then, the \gls{es} is an affine decreasing function of $CF^{r}$, and the residual scatter around the regression line in panel A is precisely the water contribution.

The weak correlation between \gls{es} and \gls{wi} follows from three compounding effects. First, the water weight is the smaller one ($W_W=0.29$). Second, and more importantly, $WI^{r}$ barely varies: a quantity that is close to zero for three-quarters of the zone set cannot correlate strongly with anything, since it hardly moves. Third, carbon intensity and water-scarcity impact are essentially decoupled across zones: their raw values are almost uncorrelated ($r=-0.03$), and so are their ratio-scaled counterparts ($r=-0.11$). The high-water-stress zones are not the high-carbon ones; they tend to be hydropower-heavy low-carbon zones (Turkey, Georgia) or moderate-carbon zones (Spain). Consequently, a zone with high \gls{wi} can sit almost anywhere on the \gls{es} axis: Turkey combines high \gls{wi} with a low \gls{es} (its carbon intensity is also high), whereas Spain combines high \gls{wi} with a high \gls{es} (its carbon intensity is low). These opposite cases cancel out and drive the correlation between \gls{es} and \gls{wi} towards zero. The resulting $r=-0.14$ is weak but correctly signed: more water stress does slightly lower the \gls{es}, but the effect is small because \gls{wi} moves little and, where it does move, it does not line up with carbon.

This near-orthogonality is not a shortcoming of the \gls{es}; it is the reason the metric is needed. If the correlation between \gls{es} and \gls{wi} were as strong as the one between \gls{es} and \gls{cf}, the water indicator would be redundant, and carbon intensity alone would already capture the water story. The fact that \gls{wi} is nearly orthogonal to both \gls{cf} and \gls{es} means that water scarcity carries information that no carbon-based metric can recover. A carbon-only scheduler is effectively blind along the \gls{wi} axis: it would treat IFCA and a non-stressed zone of the same carbon intensity as equivalent, and it would never see the summer water-stress peak of Figure~\ref{fig:aware_heatmap}. The \gls{es} is the mechanism that folds that orthogonal signal back into a single actionable scalar. The price is an explicit trade-off: because the two axes are decoupled, optimising the combined score can move carbon and water in opposite directions, exactly as observed in the DIRAC trace-driven simulation of Section~\ref{sec:dirac-trace-evaluation} (carbon down 42.8\%, water-scarcity impact up 55.6\%). Only a unified metric makes that trade-off explicit and tunable through the $W_C$ and $W_W$ weights.

Finally, it is worth noting that the \gls{es} correlates \textit{positively} with the raw volumetric \gls{wf} ($r=+0.42$): zones with a larger raw water footprint tend to obtain a \textit{better} \gls{es}, because raw \gls{wf} tracks the hydropower-rich, low-carbon Nordic generation. This confirms that the raw volumetric water footprint is not merely uninformative but actively misleading as a sustainability signal, and that the scarcity weighting embodied in \gls{wi} is indispensable.

\subsection{AI4EOSC \gls{gd} validation}
\label{sec:validation-results}

\begin{figure*}[htbp]
    \centering
    \includegraphics[width=0.8\linewidth]{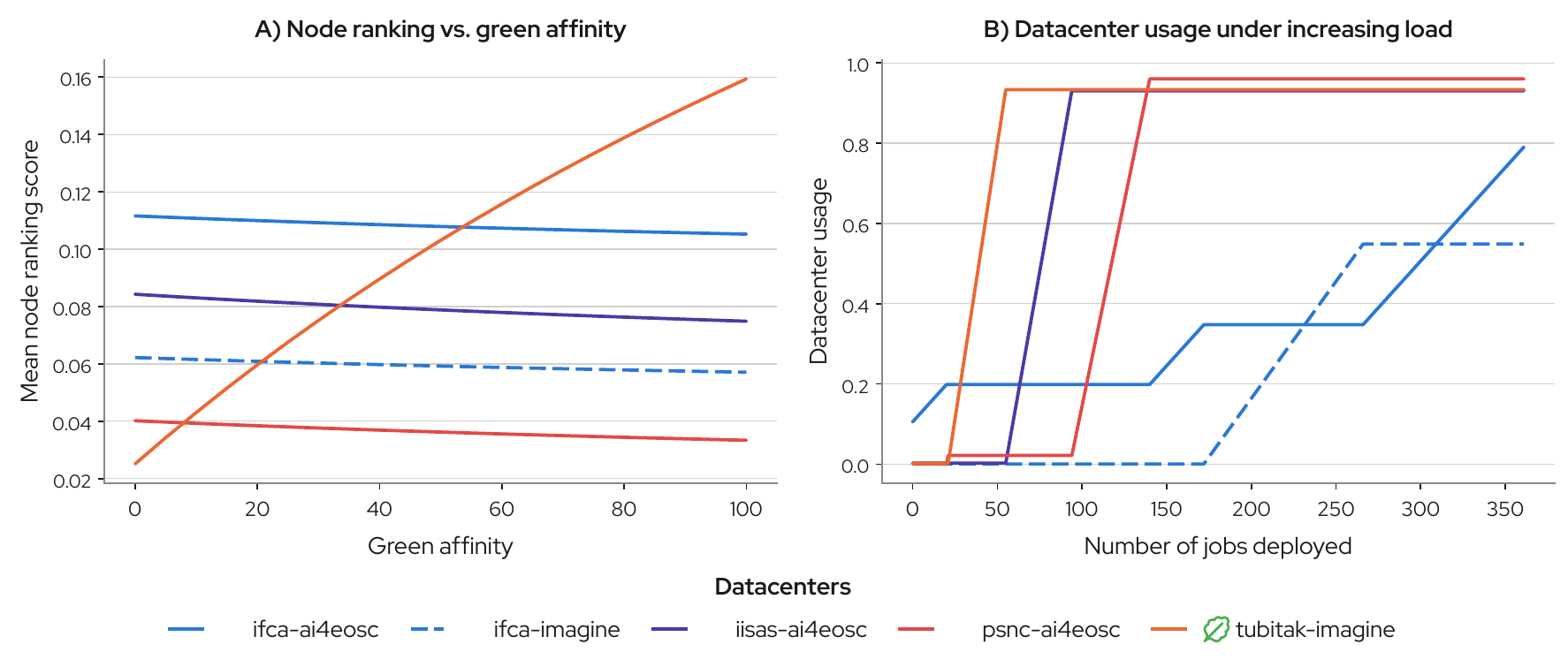}
    \caption{Results for both validation strategies, aggregating nodes in data centers. \textbf{(A)} shows the ranking computed for the different data centers, demonstrating how the node ranking scores for the same job are affected by different values of the green affinity. \textbf{(B)} demonstrates how the green data center (in green) tends to fill first when new jobs are deployed to the cluster.}
    \label{fig:results}
\end{figure*}

In Figure \ref{fig:results}, we show the results of the two validation strategies. As already mentioned, we consider the green affinity only for the \texttt{tubitak-imagine} data center, in order to better visualize the \gls{gd} effects in the scheduling.

Figure \ref{fig:results} (A) shows the Nomad ranking scores of the cluster nodes, averaged over data centers (for clarity). At zero green affinity, the ranking score is mainly dominated by the binpack score: the scoring favours allocating the job to the busiest nodes (as long as the node still has enough resources to fit the job). A minor contribution still comes from other non-green affinities (see Section \ref{sec:nomad-alloc}), but its weight is negligible with respect to binpack. As we progressively increase the value of the green affinity with the affinity score starts to take over the binpack score, resulting in the nodes belonging to the green data center (in terms of \gls{gd}) getting a higher overall ranking score.

Figure \ref{fig:results} (B) shows how the different data centers' usage increases as jobs are progressively deployed. We can clearly see that the green data center (\texttt{tubitak-imagine}) is the first one to completely fill. That does not mean it is the first one that \textit{starts to get filled}. As shown in the bottom left corner of the figure, the first 15 jobs landed in \texttt{ifca-ai4eosc}. This is likely due to the fact that the \texttt{ifca-ai4eosc} data center had, at the beginning of the experiment, already a node with a very high load. Thus, the Nomad ranker favored allocating jobs to that node with a high binpack score (so high that it completely outweighed the green affinity). Once that node is completely full, all remaining nodes have similar binpack scores, and thus the green affinity starts to make a difference, making the green data center nodes fill first. Once the green data center is filled, the two smaller data centers (with fewer and smaller nodes) are filled next. Once these are done, the remaining two larger data centers are filled progressively, the load being alternated between both.

In summary, both strategies validate the implementation of the AI4EOSC \gls{gd}. They clearly demonstrate that by effectively harnessing affinities, we are able to reduce the overall environmental impact of the AI4EOSC cluster (taking into account both carbon and water), without affecting end users nor existing platform requirements.

\subsection{DIRAC \acrfull{gsd} evaluation}
\label{sec:validation-results-dirac}

\subsubsection{Evaluation using trace-driven simulation with historically characterized sites}
\label{sec:dirac-trace-evaluation}

The controlled replay compares the randomized and \gls{gs}-based
policies using the same trace of 133,631 single-core DIRAC jobs. Both
policies use the site models, capacities, and time-aligned environmental
inputs described in Section \ref{sec:dirac-gd} and \ref{sec:dirac-validation}; only the site-selection policy
changes.

\begin{figure*}[htbp]
    \centering
    \includegraphics[width=\textwidth]
    {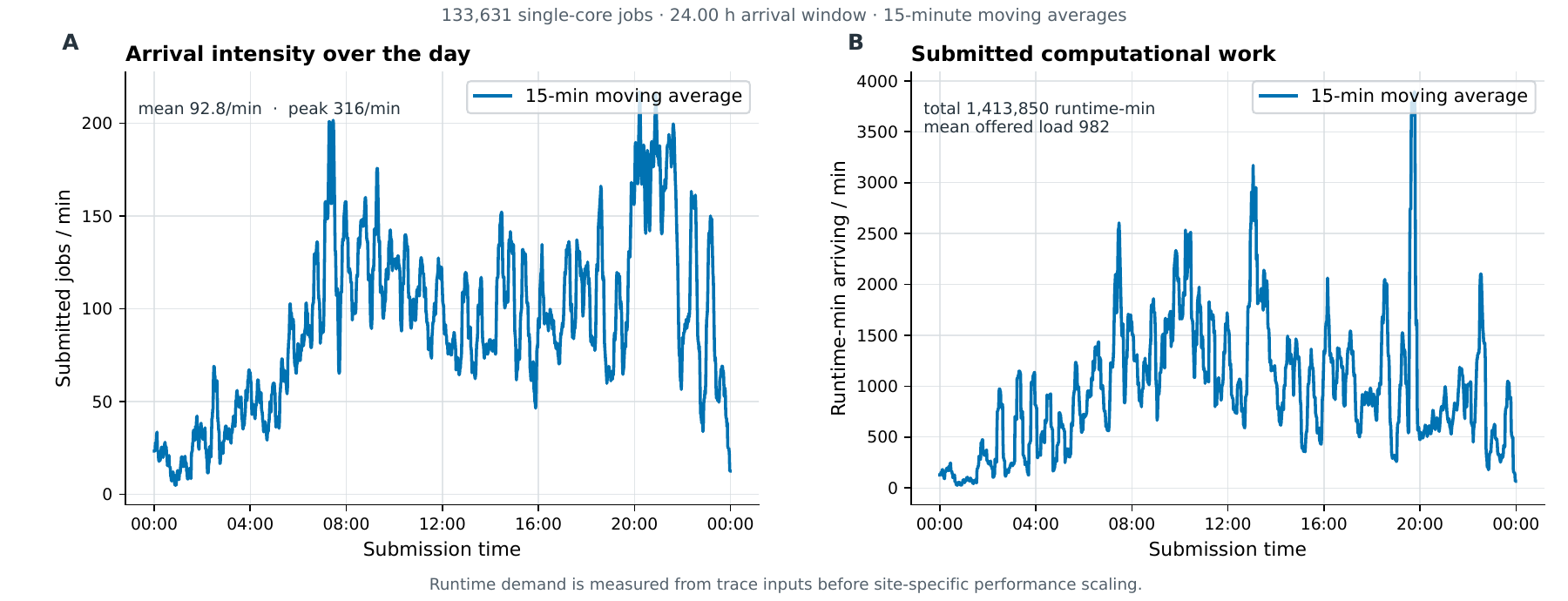}
    \caption{Temporal profile of the DIRAC workload trace used in the
simulation. Panel~A shows the 15-minute moving average of the job
submission rate. Panel~B shows the 15-minute moving average of the
submitted computational work, expressed as the sum of raw job runtime
entering the system.}
    \label{fig:workload}
\end{figure*}

Figure \ref{fig:workload} summarizes the workload submitted during the 24-hour arrival
window. Panel~A shows the number of submitted jobs per minute, whereas
Panel~B shows the corresponding computational work, expressed as the
sum of raw runtime minutes arriving in each minute. The trace contains
approximately 1.414 million runtime-minutes and
$2.120\times10^{9}$ normalized CPU-seconds.

The evaluation includes four data centers: SARA-MATRIX
(Netherlands), IN2P3-IRES (France), FZK-LCG2 (Germany), and
RAL-LCG2 (United Kingdom). The randomized policy shuffles the eligible
sites during each scheduling cycle, whereas the \gls{gs}-based policy
ranks them using the fixed historical GreenScore described in
Section~\ref{sec:dirac-gd}. Jobs remain in FIFO order under both
policies.

Table~\ref{tab:dirac-environmental-results} summarizes the complete-run
environmental results. \gls{gs}-based scheduling reduces carbon
emissions from 19.959 to 11.424~kgCO$_2$e, corresponding to a reduction
of 42.76\%. However, the AWARE-weighted water-scarcity impact increases
from 3.498 to 5.444~stress-L, corresponding to an increase of 55.64\%.

\begin{table}[H]
    \centering
    \caption{Aggregate environmental impact and computational efficiency
    for the same 133,631-job DIRAC workload under the \gls{gs}-based and
    randomized policies. Efficiency is expressed in millions of normalized
    CPU-seconds per unit of environmental impact.}
    \label{tab:dirac-environmental-results}
    \resizebox{\columnwidth}{!}{%
        \begin{tabular}{@{}lrrrr@{}}
            \toprule
            Policy &
            \shortstack{CF\\(kgCO$_2$e)} &
            \shortstack{Water\\(stress-L)} &
            \shortstack{CPU/CF\\(M CPU-s/kgCO$_2$e)} &
            \shortstack{CPU/Water\\(M CPU-s/stress-L)} \\
            \midrule
            \acrshort{gs}-based        & 11.424 & 5.444 & 185.6 & 389.5 \\
            Random  & 19.959 & 3.498 & 106.2 & 606.3 \\
            \bottomrule
        \end{tabular}%
    }
\end{table}

Because the total normalized CPU workload is identical in both runs,
a higher CPU-per-impact value indicates that the policy delivers the
same computation with a lower environmental impact. \gls{gs}-based
scheduling increases carbon efficiency by 74.71\%, but decreases
water-scarcity efficiency by 35.75\% relative to random scheduling.

\begin{figure*}[htbp]
    \centering
    \includegraphics[width=0.8\linewidth]
    {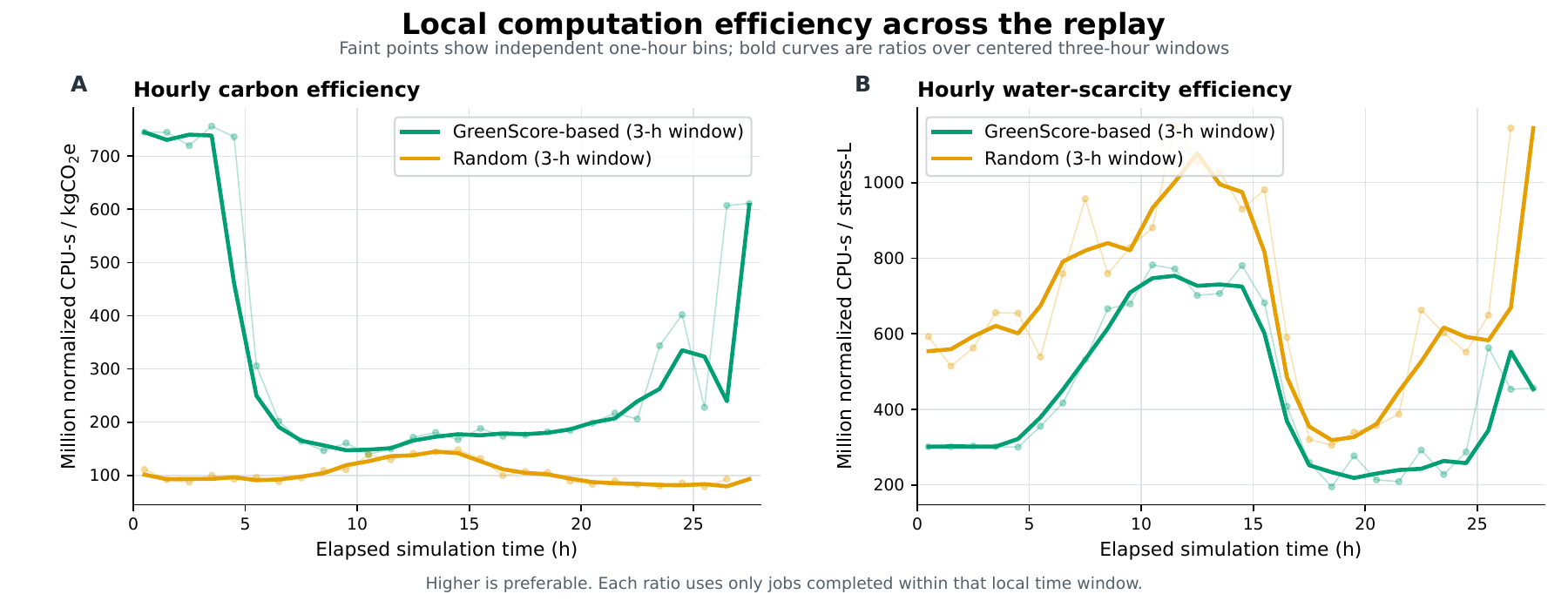}
    \caption{Environmental efficiency during the DIRAC replay. Panel A
    shows completed normalized CPU work per kgCO$_2$e. Panel B shows
    completed normalized CPU work per AWARE-weighted stress-L. Faint
    points represent independent one-hour intervals, while bold curves
    show ratios over centered three-hour windows. Higher values are
    preferable.}
    \label{fig:local-efficiency}
\end{figure*}

Figure~\ref{fig:local-efficiency} shows how these differences develop
during the replay. The quantities are non-cumulative: each point
divides the normalized CPU work completed in a local time window by
the environmental impact attributed to jobs completed in that same
window. Higher values indicate more computation per unit of impact.
Faint points represent independent one-hour intervals, while bold
curves use centered three-hour windows to reduce short-term noise.

For carbon efficiency, \gls{gs}-based scheduling outperforms random
scheduling throughout almost the entire replay. The difference is
particularly large during the first four hours, when the policy
concentrates jobs at IN2P3-IRES, the highest-ranked site under the
\gls{gs}. Carbon efficiency decreases after work begins spilling
over to additional sites, but it remains above the random baseline.

The opposite behaviour is observed for water-scarcity efficiency.
Random scheduling delivers more normalized CPU work per stress-L
during most of the replay. The \gls{gs}-based policy assigns a large
fraction of the workload to IN2P3-IRES, whose low carbon intensity
makes it favourable in the combined ranking but whose attributed
water-scarcity impact is higher during the simulation period.

Between approximately hours 10 and 15, the difference in carbon
efficiency becomes smaller as the \gls{gs}-based policy begins using
additional sites. The water-efficiency curves also vary as the
completed-job mixture, site allocation, carbon intensity, and
water-scarcity intensity change over time. Nevertheless, random
scheduling remains more water-efficient during most of the replay.

The final hourly intervals are more variable because they contain only
a small number of long-running jobs. The complete-run totals in
Table~\ref{tab:dirac-environmental-results} therefore provide the more
reliable overall comparison.

The final placement distribution explains the observed trade-off.
\gls{gs}-based scheduling assigns 84,825 jobs to IN2P3-IRES, 42,023
to SARA-MATRIX, 6,760 to RAL-LCG2, and 23 to FZK-LCG2. Random
scheduling distributes jobs more evenly, assigning 29,830 jobs to
IN2P3-IRES, 35,986 to SARA-MATRIX, 32,709 to RAL-LCG2, and 35,106 to
FZK-LCG2. By prioritizing IN2P3-IRES, \gls{gs}-based scheduling
substantially reduces carbon emissions but increases the total
water-scarcity impact.

Overall, the experiment demonstrates that \gls{gs}-based DIRAC pilot
placement improves the amount of normalized computation delivered per
unit of carbon impact while decreasing the amount delivered per unit
of water-scarcity impact. The result therefore represents a
carbon-water trade-off rather than an independent improvement in both
environmental dimensions.

\begin{figure}[t]
    \centering
    \includegraphics[width=\columnwidth]
    {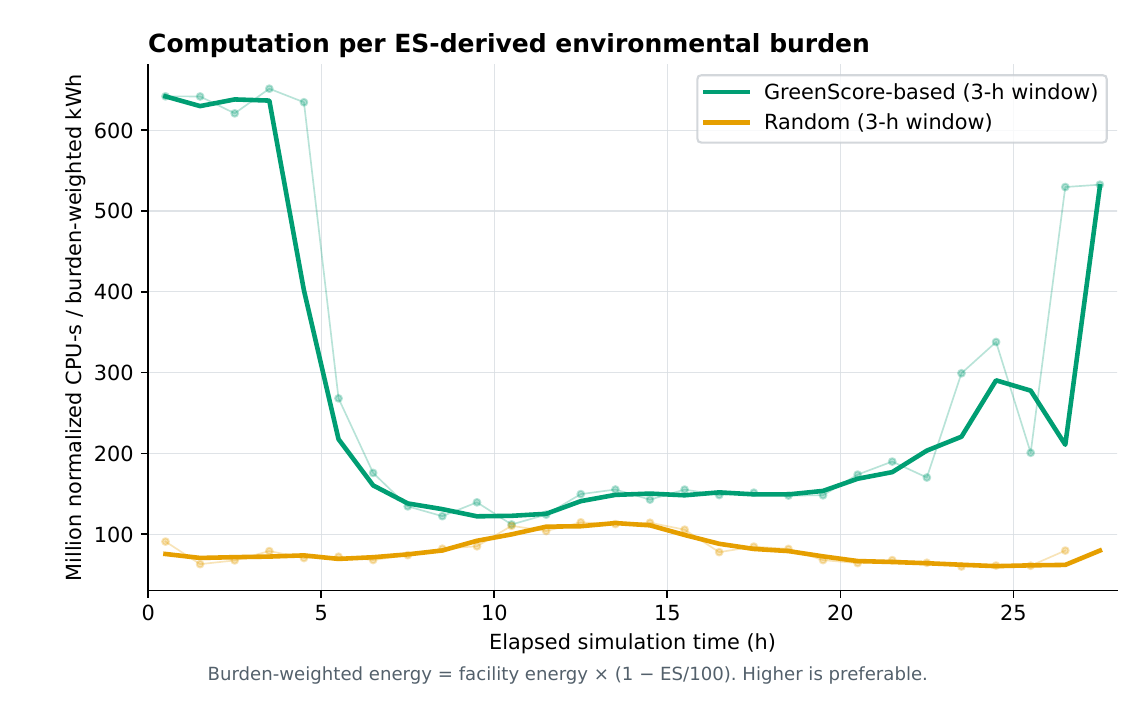}
    \caption{Normalized computation per unit of \gls{es}-derived environmental
    burden during the DIRAC replay. Faint points represent independent
    one-hour intervals, while the bold curves show ratios calculated
    over centered three-hour windows. Higher values are preferable.}
    \label{fig:dirac-environmental-burden-efficiency}
\end{figure}

To complement the separate carbon- and water-efficiency results,
Figure~\ref{fig:dirac-environmental-burden-efficiency} reports
computation relative to the combined environmental burden underlying the \gls{es}. For job $j$ executed at site $s$, this
\gls{es}-derived burden is calculated as

\begin{equation}
B_{j,s}
=
E_{j,s}\,\mathrm{PUE}_s
\left(1-\frac{\mathrm{ES}_s(t)}{100}\right),
\label{eq:dirac-es-burden}
\end{equation}

where $E_{j,s}$ is the job  energy consumption,
$\mathrm{PUE}_s$ accounts for facility-level energy overhead, and
$1-\mathrm{ES}_s(t)/100$ is the dimensionless environmental-burden
factor derived from the \gls{es}. Consequently, $B_{j,s}$
is an \gls{es} weighted facility-energy proxy, expressed in
burden-weighted kWh, rather than a physical carbon or water footprint.
Carbon and water impacts therefore remain reported separately in
kgCO$_2$e and stress-L.

Throughout most of the replay, \gls{gs}-based scheduling delivers
more normalized computation per unit of \gls{es}-derived environmental
burden than randomized scheduling. The difference is largest during
the initial period, when jobs are concentrated at IN2P3-IRES. The
advantage decreases as work spills over to additional sites, although
the \gls{gs}-based policy remains more efficient during most of the
execution. The greater variation in the final intervals results from
the small number of long-running jobs completing near the end of the
simulation.

Over the complete 133,631-job workload, \gls{gs}-based scheduling
achieves an aggregate efficiency of 156.10 million normalized
CPU-seconds per burden-weighted kWh, compared with 82.21 million for
randomized scheduling. This corresponds to an improvement of 89.88\%
in ES-derived combined environmental efficiency. This result remains
favourable despite the higher water-scarcity impact because carbon has
a larger contribution under the selected \gls{es} weighting
and normalization. It should therefore be interpreted with respect to
this combined metric, while the physical carbon and water results
remain those reported separately above.

\subsubsection{Preliminary evaluation using DIRAC deployment in production}
\label{sec:dirac-production-evaluation}

A preliminary production evaluation was conducted using jobs from the
KM3NeT \gls{vo} over two five-day observation periods. The deployment included five DIRAC sites: NIKHEF (Netherlands), SARA
(Netherlands), CNAF (Italy), CPPM (France), and IN2P3-CC (France).  One period used the
baseline randomized site ordering, while the other used
\gls{gs}-based ordering.

It is noteworthy that the production environment differs from the controlled
simulation. For instance,  pilot provisioning is opportunistic: requesting a pilot
does not guarantee that the corresponding resource will be granted or
start immediately. Pilot execution depends on site availability, local
batch queues, capacity limits, and failures. The observed placement is therefore influenced by both the
\acrfull{gsd} ordering and the dynamic state of the distributed
infrastructure.
During the \gls{gs}-based period, 53,941 completed jobs delivered
approximately $4.328\times10^8$ normalized CPU-seconds and generated
8.640~kgCO$_2$e. The random-ordering period contained 53,278 jobs,
delivered approximately $4.234\times10^8$ normalized CPU-seconds, and
generated 12.639~kgCO$_2$e. Carbon efficiency consequently increased
from 33.50 to 50.10 million normalized CPU-seconds per kgCO$_2$e,
representing an observed improvement of 49.54\%.

\begin{figure}[t]
    \centering
    \includegraphics[width=\columnwidth]
    {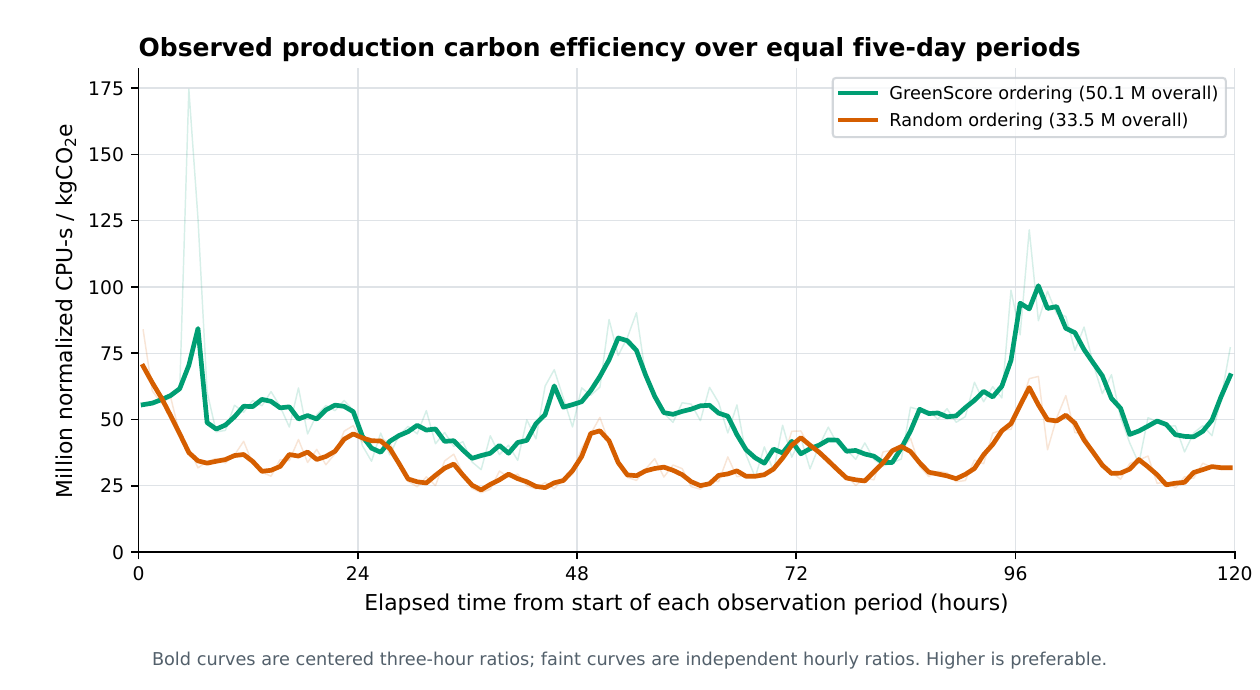}
  \caption{Production carbon efficiency during the five-day
\gls{gs}-based and randomized-ordering observations. Higher values are
preferable.}
    \label{fig:dirac-production-carbon}
\end{figure}
The corresponding AWARE-weighted water-scarcity impacts were
2.532~stress-L for \gls{gs} ordering and 2.254~stress-L for random
ordering. The differences in water-impact intensity among the five
participating sites are relatively small compared with their
differences in carbon intensity. Consequently, the effect of site
ordering is less visible for water-scarcity efficiency than for carbon
efficiency.

\begin{figure}[t]
    \centering
    \includegraphics[width=\columnwidth]
    {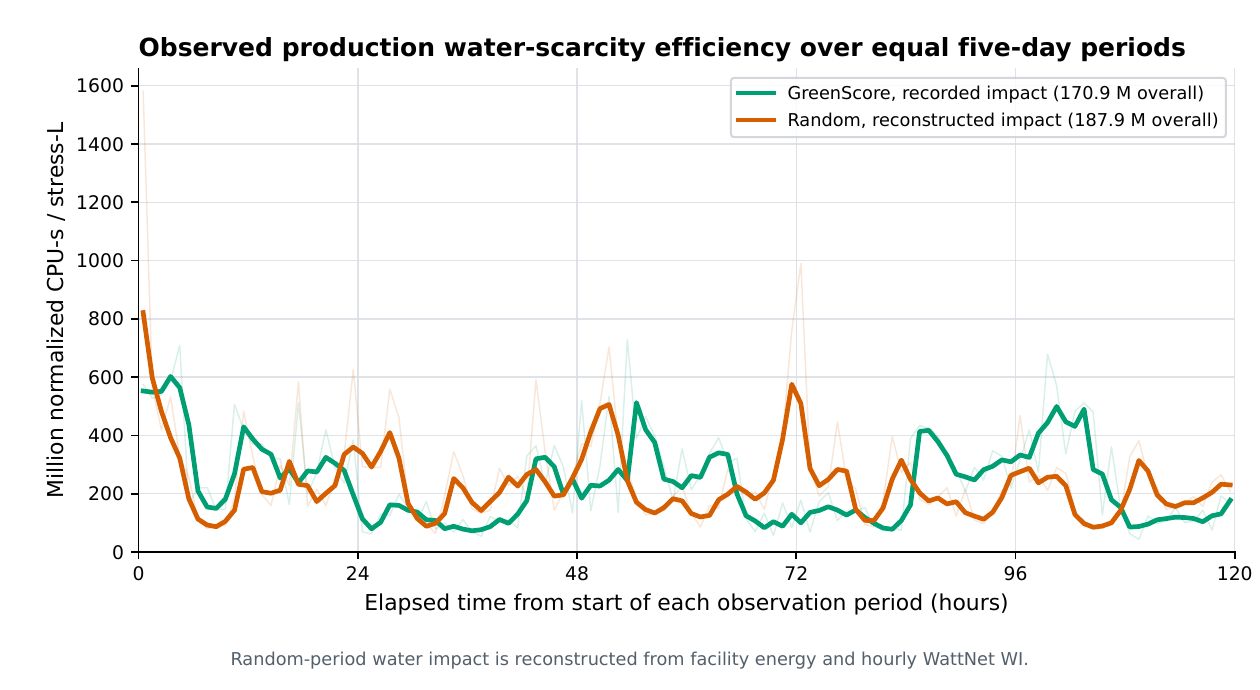}
    \caption{Production water-scarcity efficiency during the five-day
    \gls{gs} and random-ordering observations. Higher values are
    preferable.}
    \label{fig:dirac-production-water}
\end{figure}

These preliminary results show a clear difference in carbon efficiency
between the two observation periods. The effect on water-scarcity
efficiency is less pronounced because the evaluated sites have broadly
similar and relatively low water-impact intensities. Since production
pilot provisioning is opportunistic and the periods involved different
operational and environmental conditions, these results provide initial
operational evidence rather than a controlled policy comparison.

\subsection{Limitations}
\label{sec:limitations}

Several limitations of the proposed approach are worth making explicit. Collapsing the carbon and water burdens into a single index is the core of the proposal and also its main blind spot: an operator ranking sites by \gls{es} cannot tell whether a given placement trades a large amount of water for a small amount of carbon, or the reverse, without going back to the components. In the version evaluated in this work, the score combines two dimensions, carbon and water; other local externalities of electricity generation, such as land use, air quality, or nuclear waste, are not yet included. The formulation is extensible to further impact categories by design, each with its own weight, so this is a limitation of the current instantiation rather than of the approach itself. The accounting is also bounded to the operational, Scope-2-like footprint of the electricity, so on-site cooling water and the embodied impacts of hardware and facilities are out of scope, as are the assumptions and spatial-temporal resolution of the inputs the \gls{es} builds on, namely the JRC weighting set, the AWARE2.0 factors and Wattnet's flow tracing, which we take as given. Ratio scaling (Equation~\ref{scaling}) makes each value relative to the set of zones available at that instant, so a zone's score can move without any change in its own mix, which favours cross-zone comparison at a given moment over the longitudinal tracking of a single zone. Finally, the relative weight of carbon and water is held fixed, whereas the severity of the two burdens shifts with droughts, heatwaves and grid decarbonisation, and the metric has no mechanism to follow that drift.

By construction, $GS = HEE/[PUE\,(1-\mathrm{ES}/100)]$ makes hardware efficiency and grid cleanliness substitutable: a highly efficient site on a carbon-intensive grid can outrank a clean-grid site with older hardware, which can steer load away from the cleanest grids. The hardware-efficiency term itself is derived from nameplate benchmark scores and catalogue \gls{tdp} and FP32 throughput figures rather than from power and performance measured under the running workload, and the CPU/GPU energy split ($\alpha_C$, $\alpha_G$) is a provisional estimate for a typical training workload, with only two workload profiles distinguished.

The \gls{gs} optimises an intensity ratio, the useful work delivered per unit of impact, not total impact: a high-\gls{gs} site is not necessarily a low-impact one, and if its efficiency attracts a disproportionate share of the workload, its absolute footprint can still grow. More broadly, the approach decides where a workload runs, not whether it should run, be made smaller or be deferred, so the dominant driver of the problem, the absolute growth of compute demand and data-centre capacity, is outside its scope. Its value also depends on there being exploitable heterogeneity across the federation at a given moment; as grids decarbonise and converge, the headroom it exploits shrinks. It also carries no notion of saturation or fairness: it concentrates load on the greenest site until it fills, without accounting for the effect on that provider's quality of service or capacity, and if the same signal were adopted, the concentration of load widely into the same zones and windows would erode the very differentials it exploits.

Finally, the \gls{es} and \gls{gs} are predictive affinities: the evaluation shows that they steer placement, but it does not close the loop by measuring the carbon and water actually avoided against an independent accounting, so their fidelity as proxies for realised impact is assumed rather than demonstrated.

\section{Conclusion}
\label{sec:conclusion}

As computing demand is constantly growing, accelerated by resource-intensive AI workloads, mitigating the environmental impact of data centers requires moving beyond the common one-dimensional carbon accounting. Our work addressed the water-energy nexus in distributed computing by proposing the \acrlong{es}, a unified, dimensionless metric ($[0, 100]$) that allows us to jointly evaluate the \acrlong{cf} and the spatial-temporal stress-weighted \acrlong{wi} of electricity consumption. The \gls{es} integrates real-time, cross-border electricity tracking from Wattnet \cite{castrillo_melguizo_wattnet_2026}
 with monthly AWARE2.0 \cite{seitfudem_updated_2025} characterization factors, capturing the global impact of greenhouse gas emissions and the local and seasonal severity of water scarcity.

 We implemented and validated this metric in production-grade federated infrastructures through the AI4EOSC and DIRAC \glspl{gd}, the scheduling components that consume the \gls{es} to drive workload placement. In the AI4EOSC platform, our evaluation across four pan-European scientific cloud providers combined a score inspection to verify that the rankings are computed correctly, with a cluster-filling experiment to show that greener data centers are filled first. It demonstrated that incorporating \gls{es} green affinities into the workload orchestration engine steers job allocation towards the providers operating under greener mixes, in terms of both carbon and water stress, without degrading end-user experience, job scheduling latency, or hardware resource constraints.

 In the DIRAC \gls{wms}, we evaluated the \gls{gs} derived from \gls{es}, which was used for site selection both with a trace-driven simulation of 133{,}631 jobs over four sites and with a preliminary production deployment for the KM3NeT \gls{vo}. The \gls{gs}-ranked pilot placement reduced carbon emissions by 42.8\% in simulation and improved carbon efficiency by 49.5\% in production, at the cost of a higher water-scarcity impact whenever the lowest-carbon site also carried a higher local water stress. Measured against the combined \gls{es} burden, which weights carbon more heavily under the chosen normalization, the simulated run still improved environmental efficiency by roughly 90\%. This makes the carbon-water trade-off explicit and motivates tuning the \gls{es} weights to the operator's sustainability priorities.

This work advances the growing body of literature on multi-dimensional sustainable cloud computing, building upon recent efforts in spatial-temporal workload shifting \cite{attenni_spatio-temporal_2025} and carbon-water co-optimization \cite{jiang_waterwise_2025}. Our approach operationalizes standardized life-cycle assessment methodologies (ISO 14046 and AWARE2.0) directly within a production, active, and multi-tenant system. We demonstrate that water consumption from energy production can be dynamically weighted by local hydrological stress in real time, allowing the implementation of next-generation, sustainability-aware \gls{wms}. The \gls{es} is also extensible by design: further categories from the 16 environmental impacts defined by the \gls{jrc} can be added, each with its own weighting factor, to compose a metric that is progressively more representative of the overall environmental burden rather than only carbon-aware. Because the \glspl{gd} consume the \gls{es} value as provided by Wattnet, this extension is confined to the \gls{es} itself: the per-site \gls{gs} formulation and the scheduling logic require no modification and simply operate on the updated \gls{es}. Future work will explore this direction, incorporating additional impact signals, including more regional and local impacts across geographical zones, to further optimize the operational footprint of federated scientific computing.

\section*{CRediT authorship contribution statement}

\textbf{Ignacio Heredia:} Methodology, Software, Investigation, Validation, Writing -- original draft.
\textbf{Jaime Iglesias Blanco:} Methodology, Software, Data curation, Writing -- original draft, Writing -- review \& editing, Visualization.
\textbf{María Castrillo:} Conceptualization, Methodology, Formal analysis, Investigation, Validation, Writing -- original draft, Supervision.
\textbf{Andrei Tsaregorodtsev:} Conceptualization, Methodology, Validation, Investigation, Resources, Writing -- review \& editing, Supervision, Project administration, Funding acquisition.
\textbf{Mazen Ezzeddine:} Methodology, Software, Formal analysis, Investigation, Data curation, Validation, Writing -- original draft, Writing -- review \& editing, Visualization.
\textbf{Álvaro López García:} Conceptualization, Methodology, Validation, Investigation, Resources, Writing -- review \& editing, Supervision, Project administration, Funding acquisition.

\section*{Declaration of competing interest}
The authors declare that they have no known competing financial interests or personal relationships that could have appeared to influence the work reported in this paper.

\section*{Acknowledgments}
The authors acknowledge the funding and support from: the AI4EOSC project ``Artificial Intelligence for the European Open Science Cloud'', which has received funding from the European Union's Horizon Europe research and innovation programme under grant agreement number 101058593; the GreenDIGIT project ``Greener Future Digital Research Infrastructures'', which has received funding from the European Union's Horizon Europe research and innovation programme under grant agreement number 101131207. ALG and MC also acknowledge the support from the \textit{Consejería de Educación, Formación profesional y Universidades} of the \textit{Gobierno de Cantabria} via the ``\textit{Actividad estructural para el desarrollo de la investigación del Instituto de Física de Cantabria}'' project.

\section*{Data and code availability}
The environmental data used in this study, namely the \gls{cf}, \gls{wf}, \gls{wi}, and \gls{es} at 15-minute resolution for the European electricity zones, are not redistributed as part of this work. Instead, these data are obtained directly from the Wattnet service through its public API (\url{https://api.wattnet.eu/})~\cite{castrillo_melguizo_wattnet_2026}, ensuring that the experiments rely on the same data source and methodology used to provide these environmental indicators. The AWARE2.0 water-scarcity characterization factors used to derive the corresponding water-related impacts are those published in~\cite{seitfudem_updated_2025, georg_seitfudem_2025_16332127}. These factors are applied consistently with the methodology described in the referenced studies and are not redistributed separately with the software or experimental datasets.

The \gls{gs}-based scheduling logic developed in this work is available as open-source software through two implementations: the DIRAC \gls{gsd}, available at \url{https://github.com/GreenDIGIT-project/GreenDIRAC}, which extends DIRAC Interware (GPLv3)~\cite{diracgrid_github}; and the AI4EOSC \gls{gd}, integrated into the AI4EOSC Platform API and available at \url{https://github.com/ai4os/ai4-papi} (Apache-2.0)~\cite{HEREDIA2026108672}.

\bibliographystyle{elsarticle-num-names}
\bibliography{references.bib}

\end{document}